\documentclass[journal=./achemso/jpc,manuscript=article,layout=onecolumn]{achemso}
\usepackage{natmove}
\usepackage{physics}
\usepackage{amsmath}
\usepackage{graphicx}
\usepackage{mdframed}
\usepackage{color,soul}
\usepackage{appendix}
\usepackage{mathtools}  
\usepackage{amssymb}  
\usepackage{natbib}
\usepackage{float}
\setkeys{acs}{usetitle = true}

\SectionNumbersOn

\title{Tensor-Train Split-Operator KSL (TT-SOKSL) Method for Quantum Dynamics Simulations} 

\author{Ningyi Lyu}
\affiliation{Department of Chemistry, Yale University, P.O. Box 208107, New
Haven, CT 06520-8107} 
\author{Micheline B. Soley}
\affiliation{Department of Chemistry, Yale University, P.O. Box 208107, New
Haven, CT 06520-8107} 
\alsoaffiliation{Yale Quantum Institute, Yale University, P.O. Box  208334, New Haven, CT 06520-8263} 
\author{Victor S. Batista}
\affiliation{Department of Chemistry, Yale University, P.O. Box 208107, New
Haven, CT 06520-8107} 
\alsoaffiliation{Yale Quantum Institute, Yale University, P.O. Box  208334, New Haven, CT 06520-8263} 
\email{victor.batista@yale.edu}

\makeatother

\begin{document}
\begin{abstract}
Numerically exact simulations of quantum reaction dynamics, including non-adiabatic effects in excited electronic states, are essential to gain fundamental insights into ultrafast chemical reactivity and rigorous interpretations of molecular spectroscopy. 
Here, we introduce the tensor-train split-operator KSL (TT-SOKSL) method for quantum simulations in tensor-train (TT)/matrix product state (MPS) representations. TT-SOKSL propagates the quantum state as a tensor train using the Trotter expansion of the time-evolution operator, as in the tensor-train split-operator Fourier transform (TT-SOFT) method. However, the exponential operators of the Trotter expansion are applied using a rank adaptive TT-KSL scheme instead of using the scaling and squaring approach as in TT-SOFT. We demonstrate the accuracy and efficiency of TT-SOKSL as applied to simulations of the photoisomerization of the retinal chromophore in rhodopsin, including non-adiabatic dynamics at a conical intersection of potential energy surfaces. The quantum evolution is described in full dimensionality by a time-dependent wavepacket evolving according to a two-state 25-dimensional model Hamiltonian. We find that TT-SOKSL converges faster than TT-SOFT with respect to the maximally allowed memory requirement of the tensor-train representation and better preserves the norm of the time-evolving state. When compared to the corresponding simulations based on the TT-KSL method, TT-SOKSL has the advantage of avoiding the need of constructing the matrix product state Laplacian by exploiting the linear scaling of multidimensional tensor train Fourier transforms.  
\end{abstract}

\section{Introduction}
Simulations of quantum phenomena in chemical and biological systems\cite{Cao2020,Marais2018} typically require time-dependent methods. For example, photoinduced reactions, \cite{Domcke2012,Nelson2020,Nelson2014,Zhang2020} as well as processes that involve energy transfer,\cite{Mulvihill2021} electron transfer,\cite{Yamijala2021,Brian2021,Tong2020,Sato2018,Marmolejo-Valencia2021} simulations of molecular spectroscopy,\cite{Yan1988} and coherent control,\cite{Rego2009} require rigorous descriptions of quantum effects, including tunneling, interference, entanglement, and non-adiabatic dynamics.\cite{Baer2006,Yarkony2012}
Simulations in the time-dependent picture require integration of the time-dependent Schr\"odinger equation (TDSE) explicitly, which can be efficiently performed for small molecular systems for example by using the split-operator Fourier transform (SOFT) method , which is a numerically exact method.~\cite{Feit1982,Feit1982a,Kosloff1983} However, SOFT is limited to systems with very few degrees of freedom (DOF) ({\em i.e.}, molecular systems with less than 4 or 6 atoms),~\cite{Meyer09, Nyman13} since it is based on a full basis set representation requiring storage space and computational effort that scale exponentially with the number of coupled DOFs. Utilizing an adaptive grid that evolves simultaneously with the wavepacket, the capability of SOFT is extended to successfully treat the dynamics of an eight-dimensional Henon-Heiles model.\cite{Choi2019} Other numerically exact quantum dynamics methods include the Chebyshev polynomial expansion method\cite{Kosloff1983, Chen1999, Nyman13} and methods based on the Krylov expansion.\cite{Sidje1998}

The exponential scaling problem has motivated the development of a variety of methods based on truncated basis sets. Some of these methods employ Gaussian coherent states, such as the method of coupled coherent states\cite{Shalashilin04, Shalashilin08}, the multiple-spawning method \cite{Ben-Nun02, Yang09}, and the matching-pursuit algorithm.\cite{Wu03} The Multi-Configurational Time-Dependent Hartree (MCTDH) method groups DOFs into ``particles'' represented in a DVR basis\cite{Meyer90, Beck00, Meyer03, Worth08, Meyer09} and has been implemented for efficient calculations in terms of the so-called multilayer MCTDH method.\cite{Wang15} However, determining how exactly the basis should be truncated or how to group DOFs into particles can be difficult and relies on approximations.\cite{Worth08} It is also noted that the MCTDH equations of motion involve ill-conditioned matrices which require smaller time steps at the beginning of the propagation \cite{Lubich2014a}, or special stepsize adaptive techniques.\cite{Lindoy2021}

In earlier work, we have introduced the so-called tensor-train (TT) split-operator Fourier transform (SOFT) method (TT-SOFT)\cite{Greene2017} that allows for rigorous simulations of multidimensional nonadiabatic quantum dynamics. TT-SOFT represents the time-dependent wave function as a dynamically adaptive TT and evolves it by recursively applying the time-evolution operator as defined by the Trotter expansion. Exploiting the efficient TT implementation of multidimensional Fourier transforms, the TT-SOFT algorithm applies the Trotter expansion of the time-evolution operator using exponential operators in tensor-train format generated by the scaling and squaring method. The accuracy and efficiency of the TT-SOFT method were demonstrated as applied to the propagation of 24-dimensional wave packets describing the $S_1$/$S_2$ non-adiabatic dynamics of interconversion of pyrazine after UV photoabsorption. Here, we introduce the tensor-train split-operator KSL (TT-SOKSL) method, which, although it is also based on the Trotter expansion of the time-evolution operator, has the advantage of avoiding the computational bottleneck of scaling and squaring by applying the exponential operators according to a rank-adaptive version of the TT-KSL solver\cite{Lubich2015,Lubich2014,Koch2007}.

The TT-KSL algorithm\cite{Lubich2015,Lubich2014} is a TT implementation of the dynamical low-rank approximation (DLRA) method for evolution of time-dependent matrices, where the name KSL comes from the DLRA integration scheme that successively updates three component matrices $\mathbf{K}, \mathbf{S}$ and $\mathbf{L}$.\cite{Koch2007} Rather than generating a low-rank approximation by rounding after generating a high-rank solution, TT-KSL implements an orthogonal projection onto a low-rank manifold according to the Dirac-Frenkel Time-Dependent Variational Principle\cite{Dirac1930,Frenkel1934,McLachlan1964,Heller1976} (TDVP). The capability of TT-KSL is demonstrated in applications to simulations of pyrazine \cite{Xie2019}, the Fenna-Matthews-Olson (FMO) complex \cite{Li2020}, and singlet fission in molecular dimer and Perylene-Bisimide aggregates.\cite{Baiardi2019} The method has typically been carried out as an effective matrix-vector multiplication scheme in the occupation number representation using the kinetic operator matrix in TT format. When implemented in the position grid representation, the kinetic operator requires a finite-difference approximation\cite{kazeev2012low} or implementations based on the Fourier Grid Hamiltonian (FGH)\cite{ClayMarston1989,Balint-Kurti1992,Stare2003} or similar Discrete Variable Representation (DVR) methods.\cite{Dickinson1968,Colbert1992} The TT-SOKSL method avoids matrix-vector multiplication schemes by exploiting the diagonal representation of the kinetic operator in momentum space to evolve the quantum state by elementwise vector-vector multiplication. TT-SOKSL thus combines the simplicity of TT-SOFT and the advantages of a projector-splitting integrator to implement the exponential operators of the Trotter expansion in diagonal form.

We demonstrate the capabilities of TT-SOKSL as applied to simulations of non-adiabatic quantum dynamics. We focus on the photoisomerization process of the retinal chromophore in rhodopsin as described by a two-state 25-mode model Hamiltonian\cite{Hahn2000}. The model system is ideally suited for comparisons to calculations based on TT-SOFT and the MCTDH methods. 


\section{Methods}
\subsection{Potential energy surface}
\begin{figure*}
\includegraphics[scale=0.4]{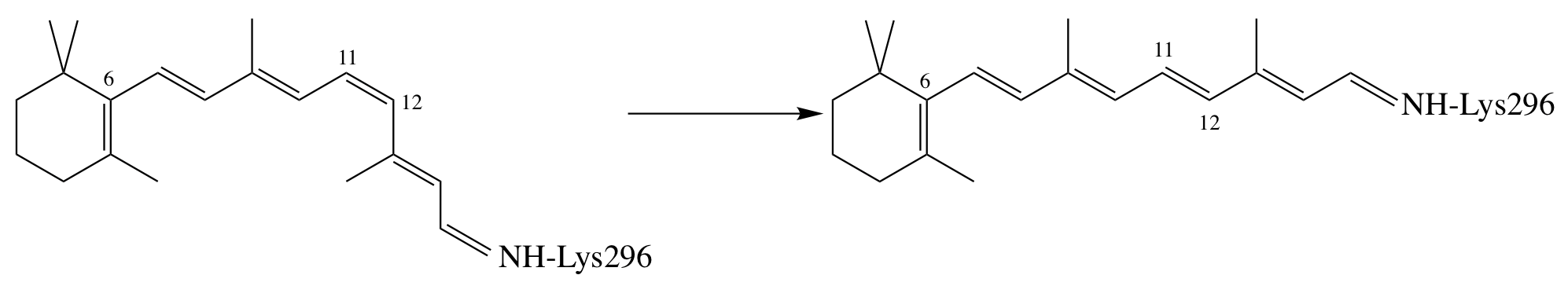}
\caption{11-cis/all-trans photoinduced isomerization of the retinyl chromophore in visual rhodopsin.}
\label{fig:lewis}
\end{figure*}

\begin{figure*}
\includegraphics[scale=0.4]{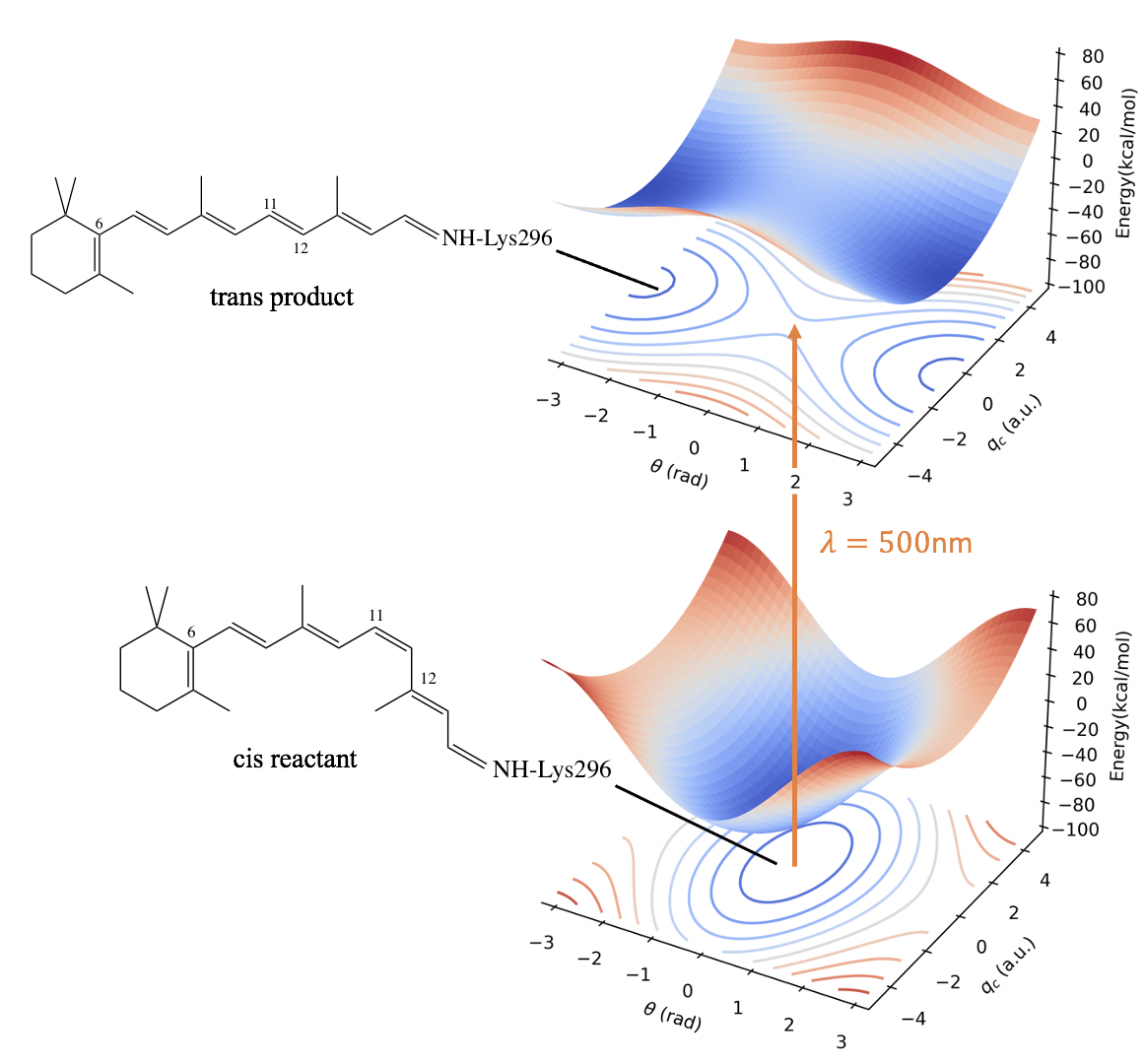}
\caption{Diabatic potential energy surfaces of the rhodopsin ground ($S_0$) and first excited ($S_1$) electronic states as a function of the two large amplitude coordinates, the torsion dihedral angle coordinate $\theta$ of the $\text{C}_{11}\text{C}_{12}$ bond and the stretching coordinate $q_c$ of the conjugated polyene chain. Black lines connect cis/trans wells with the cis/trans conformations of the molecule. The orange arrow indicates the initial photo-excitation ($\lambda=500$~nm) to the $S_1$ surface.}
\label{fig:PESs}
\end{figure*}

We simulate the nuclear and electronic dynamics of the photoisomerization of the retinal molecule (Figure \ref{fig:lewis}) to explore the capabilities of the TT-SOKSL method as compared to TT-SOFT and MCTDH.
The model Hamiltonian \cite{Hahn2000,Hahn2000a} consists of the vibronically coupled $S_0$ and $S_1$ diabatic potential energy surfaces (PES). These are 25-dimensional PESs parametrized by the resonance Raman active modes of the retinyl chromophore in rhodopsin. Two modes are identified as the large-amplitude primary modes, which correspond to the $\text{C}_{11}$=$\text{C}_{12}$ torsion and the ethylenic stretching of the polyene chain (Figure \ref{fig:PESs}). 
The other $23$ modes are modeled as harmonic oscillators linearly coupled to the excited electronic state with frequencies and equilibrium positions parametrized by the experimental resonance Raman spectrum, as follows:
\begin{equation}\label{retiH}
\begin{split}
&H=H_S+H_B,\\
&H_S=\sum_{n,m=0}^1|\psi_n^{el}\rangle (T_s\delta_{mn}+V_{mn})\langle \psi_n^{el}|,\\
&T_s=-\frac{\Omega_c}{2}\frac{\partial ^2}{\partial q_c^2}-\frac{1}{2I}\frac{\partial^2}{\partial \theta^2},\\
&V_{00}=\frac{W_0}{2}(1-\text{cos}\theta)+\frac{\Omega_c}{2}q_c^2,\\
&V_{11}=E_1-\frac{W_1}{2}(1-\text{cos}\theta)+\frac{\Omega_c}{2}q_c^2+\kappa q_c,\\
&V_{10}=V_{01}=\lambda q_c,\\
&H_B=\sum_{n=0}^1\sum_{i=1}^{23}|\psi_n^{el}\rangle (\frac{\omega_i}{2}(\frac{\partial ^2}{\partial q_c^2}+q_i^2)+\kappa _iq_i\delta_{n1})\langle \psi_n^{el}|.
\end{split}
\end{equation}
Here, $\Omega_c=1532~\text{cm}^{-1}$, with remaining parameters (in eV): $1/I=4.84\times 10^{-4}, E_1=2.48, W_0=3.6, W_1=1.09, \kappa=0.1, \lambda=0.19$.\cite{Hahn2000} Parameters of the other vibrational modes $\{\kappa_i, \omega_i\}$ as well as our Python codes for reproducing all of the results reported in this paper are available at https://github.com/NingyiLyu/TTSOKSL. We note that a new set of parameters has been recently reported \cite{Johnson2017} to better match experimental findings\cite{Johnson2015,Johnson2017}. Nevertheless, our calculations are based on the original set of parameters to allow for comparisons to earlier studies. \cite{Balzer2004,Chen2007,Videla2018,Axelrod2019,flores2004model}

\subsection{Initial Conditions}
We initialize the wavepacket as a Gaussian on the $S_1$ excited state, centered at the equilibrium position of the vibrational modes in the ground electronic state, to simulate the light-induced $\text{S}_1 \leftarrow \text{S}_0$ vertical transition:
\begin{equation}\label{psi0}
\psi(x_1,...,x_{25} ;0)=\prod_{j=1}^{25}(\sigma_j^2\pi/2)^{-1/4}e^{-x_j^2/\sigma_j^2},
\end{equation}
where $\sigma_1=0.15228$ a.u., $\sigma_j=\sqrt{2}$ a.u., with $j=2$--$25$. 
The bath mode width parameters $\sigma_i=\sqrt{2}\text{ a.u.}$ have been defined according to the model presented in ref.~\cite{Hahn2000} in place of the standard harmonic fit to the electronic ground potential energy surface at the equilibrium geometry in order to facilitate direct comparison to literature results.  The wavepacket then evolves according to the coupled $S_1/S_0$ potential energy surfaces, which partitions the population between the cis/trans conformations in the ground and excited states as it reaches configurations close to the conical intersection. The isomerization quantum yield ({\em i.e.}, trans:trans+cis population ratio) is computed by integrating the trans population as defined by configurations with $|\theta|>\pi/2$ where, according to Eq.~\eqref{retiH}, $\theta$ is the dihedral angle about the $\text{C}_{11}$=$\text{C}_{12}$ bond. 

\subsection{Tensor-train decomposition}
TT-SOKSL relies on the tensor-train (TT) format,\cite{Oseledets2011,Oseledets2010,Grasedyck2009,Hackbusch2009} also called matrix product states (MPS) with open boundary conditions,\cite{Stlund1995,Verstraete2008,Orus2014,Paeckel2019,Larsson2019} recently explored for the development of methods for quantum dynamics and global optimization.\cite{Greene2017,soley2021iterative,soley2021functional} 

The TT format of an arbitrary $d$-dimensional tensor \(X\in \mathbb{C}^{n_1\times...\times n_d}\) involves a train-like matrix product of $d$ 3-dimensional tensors \(X_i\in \mathbb{C}^{r_{i-1}\times n_i\times r_i}\) with \(r_0=r_d=1\), so any element $X(j_1,...,j_d)$ of \(X\) can be evaluated, as follows:\cite{Oseledets2011}
\begin{equation}\label{TT}
X(j_1,...,j_d)=\sum_{a_0=1}^{r_0} \sum_{a_1=1}^{r_1}...\sum_{a_{d}=1}^{r_{d}}X_1(a_0,j_1,a_1)X_2(a_1,j_2,a_2)...X_d(a_{d-1},j_d,a_d),
\end{equation}
or equivalently in matrix product notation,
\begin{equation}\label{compact}
X(j_1,...,j_d)=\mathbf{X_1}(j_1)...\mathbf{X_d}(j_d),
\end{equation}
where \(\mathbf{X_i}(j_i)\in \mathbb{C}^{r_{i-1}\times r_i}\) is the \(j_i\)-th slice of tensor core \(X_i\). Throughout this paper, bold capital letters are used to denote matrices, while Italian capital letter are used to denote multi-dimensional tensors. The TT-rank (referred as the bond dimension in the theoretical physics community\cite{Larsson2019}) is defined in terms of the vector $r=\{r_0,r_1,...,r_{d-1},r_d\}$ introduced by Eq.~(\ref{TT}). We note that the resulting TT representation of $X$, with $n_1=n_2=\dots=n_d=n$ and $r_1=r_2=\dots=r_{d-1}=\tilde{r}$, requires only $n \times d \times \tilde{r}^2$ data points, which bypasses the usual exponential $n^d$ number of elements required by the full-dimensional representation. 

The TT format of operator $M\in \mathbb{C}^{(n_1\times n_1)\times (n_2\times n_2)\times...\times (n_d\times n_d)}$
, referred as TT matrix or matrix-product-operator (MPO), is defined with analogous format, 
\begin{equation}\label{TTmatdef}
    M(l_1,\dots,l_d; j_1,\dots,j_d)=\mathbf{M}_1(l_1,j_1)\dots\mathbf{M}_d(l_d,j_d),\quad~\text{with}\quad~\mathbf{M}_i(l_i,j_i)\in \mathbb{C}^{r_{M_{i-1}}\times r_{M_i}}.
\end{equation}
Therefore, TT-matrices $M$ operate core-by-core on TT vectors $X\in \mathbb{C}^{n_1\times...\times n_d}$, as follows:
\begin{equation}\label{TTmat}
\begin{split}
&MX(l_1,\dots,l_d)=\sum_{j_1,\dots,j_d}M(l_1,\dots,l_d;j_1,\dots,j_d)X(j_1,\dots,j_d),\\
&\;\;\;\;\;\;\;\;\;\;\;\;\;\;\;\;\;\;\;\;\;\;\;\;=\sum_{j_1,\dots,j_d}(\mathbf{M}_1(l_1,j_1)\otimes \mathbf{X}_1(j_1))\dots(\mathbf{M}_d(l_d,j_d)\otimes \mathbf{X}_d(j_d)).
\end{split}
\end{equation}

\subsection{TT-SOKSL}
\subsubsection{Split-operator propagator}
The TT-SOKSL method integrates the time-dependent Schr\"odinger equation:
\begin{equation}
\dot{\Psi}(t)=-\frac{i}{\hbar}\hat{H}\Psi(t)
\label{eq:schr}
\end{equation}
where $\hat{H}=\hat{T}+\hat{V}$ is the Hamiltonian of the system and $\Psi(t)$ is the time-dependent state. 
Equation~(\ref{eq:schr}) can be integrated to second-order accuracy by using the Trotter (Strang-splitting) approximation, as in the SOFT method.~\cite{Feit1982,Feit1982a,Kosloff1983} For each integration time-step $\tau=t_{k+1}-t_k$, we evolve the state from $\Psi(t_k)$ to $\Psi(t_{k+1})=\Psi(t_k+\tau)$, as follows. First, we obtain $\Psi_1(t_k+\tau)$ by integrating the equation,
\begin{equation}
\label{eq:SOKSLsplit1}
\begin{split}
&\dot{\Psi}_1(t)=-\frac{i}{2\hbar}\hat{V}{\Psi}_1(t),
\end{split}
\end{equation}
with ${\Psi}_1(t_k)=\Psi(t_k)$. Then, we obtain $\Psi_2(t_k+\tau)$ by integrating the equation,
\begin{equation}
\label{eq:SOKSLsplit2}
\begin{split}
&\dot{\Psi}_2(t)=-\frac{i}{\hbar}\hat{T}{\Psi}_2(t),
\end{split}
\end{equation}
with ${\Psi}_2(t_k)={\Psi}_1(t_k+\tau)$ and we obtain $\Psi_3(t_k+\tau)$ by integrating the equation,
\begin{equation}
\label{eq:SOKSLsplit3}
\begin{split}
&\dot{\Psi}_3(t)=-\frac{i}{2\hbar}\hat{V}{\Psi}_3(t),
\end{split}
\end{equation}
with ${\Psi}_3(t_k)=\Psi_2(t_k+\tau)$ to obtain $\Psi(t_{k+1}) = \Psi_3(t_k+\tau)+O(\tau^3)$. 

TT-SOKSL represents $\Psi(t)$ as a TT vector and sequentially integrates Eqs.~\eqref{eq:SOKSLsplit1}--\eqref{eq:SOKSLsplit3} by using the rank-adaptive implementation of the so-called dynamical low-rank approximation
(TT-KSL method, Appendix~\ref{sec:ttksl}).\cite{Lubich2015,Lubich2009} Our rank-adaptive scheme ensures that the TT rank does not limit the accuracy of the propagation. As in the TT-SOFT algorithm, TT-SOKSL exploits the efficient implementation of multidimensional Fourier transforms in TT format. So, Eq.~(\ref{eq:SOKSLsplit1}) is integrated in the coordinate representation by elementwise vector-vector multiplication (equivalent to diagonal matrix-vector multiplication), using the TT-vector operator $-i\hat{V}/2\hbar$. $\Psi_1(t_k+\tau)$ obtained in TT format is then Fourier transformed (FT), and Eq.~(\ref{eq:SOKSLsplit2}) is integrated in momentum space also by vector-vector multiplication using the TT vector $-i\hat{T}/\hbar$. After an inverse Fourier Transform (IFT) of $\Psi_2(t_k+\tau)$ back to the coordinate representation, Eq.~(\ref{eq:SOKSLsplit3}) is solved akin to Eq.~(\ref{eq:SOKSLsplit1}) to obtain the time-evolved state, as follows: 
\begin{equation}\label{TTSOKSLmain}
\begin{split}
&\Psi({\bf x},t_{k+1})= \\
&\text{KSL}\left(\tau,-i\hat{V}({\bf x})/2\hbar,  \text{ FT}\left[ \text{KSL}\left(\tau, -i \hat{T}({\bf p})/ \hbar, \text{ IFT}\left[\text{KSL}\left(\tau, -i\hat{V}({\bf x}')/2\hbar, \Psi({\bf x}',t_k)   \right)\right]\right)\right]\right),\\
\end{split}
\end{equation}
where $\text{KSL}$ denotes the rank-adaptive TT-KSL integration substeps. 

Eq.~\eqref{TTSOKSLmain} shows the relationship between TT-SOKSL and TT-KSL. In TT-SOKSL, the potential and kinetic energy operators are applied as diagonal TT matrices in the coordinate and momentum representations, respectively. In contrast, TT-KSL represents the Hamiltonian as a dense TT matrix. TT-SOKSL therefore diverges from TT-KSL in that it bypasses the need to construct dense matrix product operators in TT format as required by the original TT-KSL method. 

Each of the Eqs.~\eqref{eq:SOKSLsplit1}--\eqref{eq:SOKSLsplit3} can be represented, as follows:
\begin{equation}
\dot{\Psi}(t) = {\bf M} \Psi(t),
\label{eq:shr1}
\end{equation}
where ${\bf M}$ is the {\em diagonal} TT-matrix representation of a TT vector, defined as ${\bf M} = -i{\bf V}/2\hbar$ for Eqs.~\eqref{eq:SOKSLsplit1} and \eqref{eq:SOKSLsplit3} and ${\bf M}=-i{\bf T}/\hbar$ for Eq.~\eqref{eq:SOKSLsplit2}, with ${\bf V}$ and {\bf T} the potential and kinetic energy matrices in coordinate- and momentum-space representations, respectively.  

\subsubsection{Dynamical Low-Rank Approximation: KSL algorithm}
\begin{figure}[H]
\includegraphics[scale=0.3]{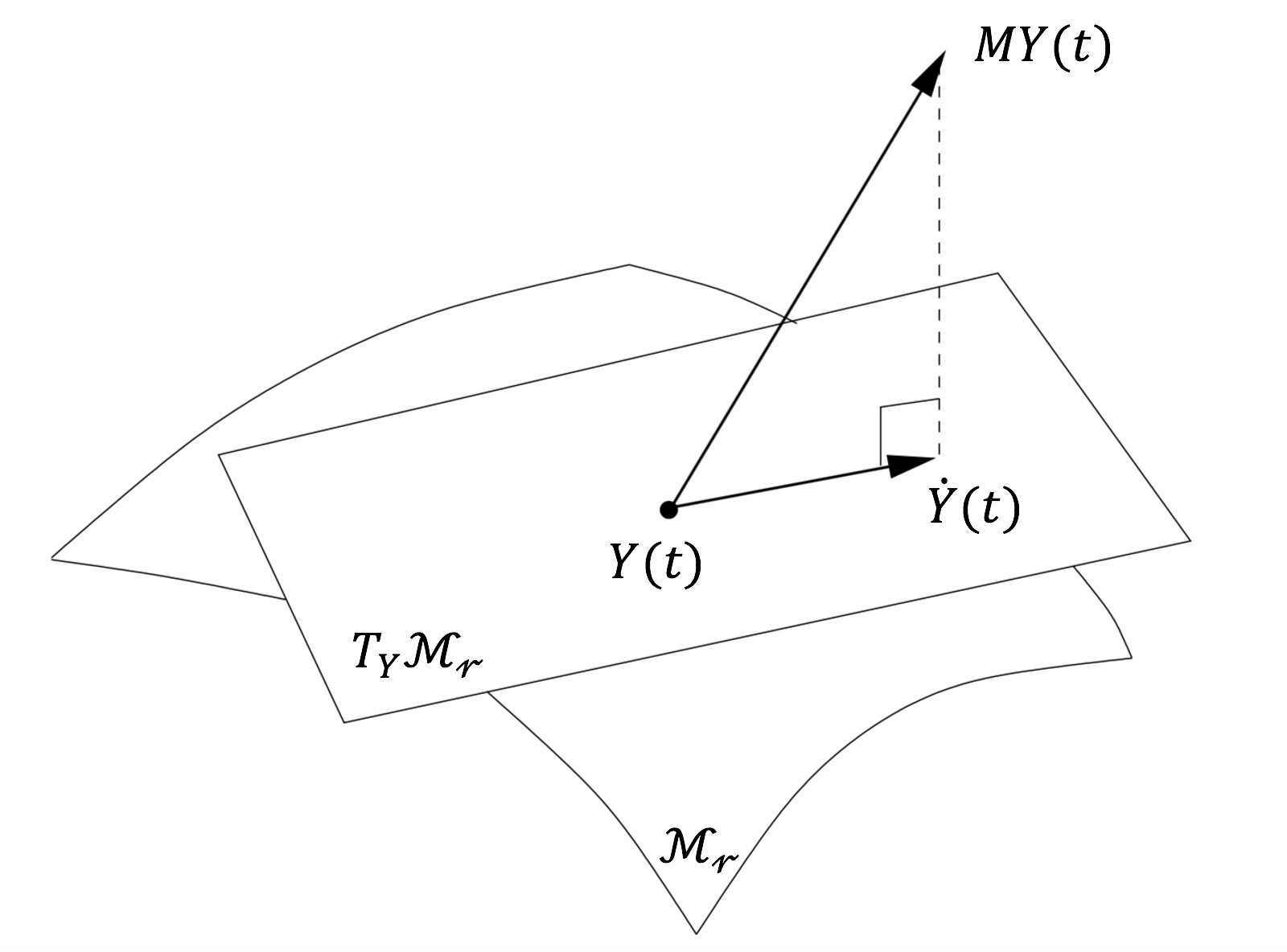}
\caption{Representation of the orthogonal projection of \(MY(t)\) onto the plane tangent to the manifold \(\mathcal{M}_r\) at \(Y(t)\).
Adapted from Ref.~\citealp{Lubich2009}.}
\label{fig:TDVP}
\end{figure}

The dynamical low-rank approximation method\cite{Lubich2009} is an efficient algorithm to obtain an approximate solution of Eq.~(\ref{eq:shr1}) in the form of a matrix \(\mathbf{Y}(t)\) of specified rank~\(r\). Rather than obtaining a solution with high rank and then truncating it by singular value decomposition (SVD), the dynamical low-rank approximation method integrates the following equation,
\begin{equation}\label{DLRA}
\dot{\mathbf{Y}}(t)=P_\mathbf{Y}({\bf M} \mathbf{Y}(t)),
\end{equation}
The operator $P_\mathbf{Y}$ projects \({\bf M}\mathbf{Y}(t)\) onto the tangent plane \(T_\mathbf{Y} \mathcal{M}_r\) (Figure~\ref{fig:TDVP}) --{\em i.e.}, the plane tangent to the manifold \(\mathcal{M}_r\) of states of rank $r$ at \(\mathbf{Y}(t)\).  
After each propagation time-step $\tau$, the resulting approximate solution $\mathbf{Y}(t+\tau)$ is the state on the manifold \(\mathcal{M}_r\) that is closest to the exact higher-rank solution.

Appendices A--D provide the derivation of $P_\mathbf{Y}$ and its implementation according to Eq.~(\ref{DLRA}). An important advantage of the KSL propagation scheme is that it does not require matrix inversion or any kind of regularization scheme as typically implemented in other propagation methods, such as MCTDH.\cite{regul}

The TT-KSL method \cite{Lubich2015} is the tensor-train implementation of the dynamical low-rank approximation. It is based on
the following expression of $P_Y$ (with derivation and implementation explained in Appendices E-F):
\begin{equation}\label{PXZmain}
P_Y(MY)=\sum_{i=1}^{d-1}\text{Ten}_i[(\mathbf{I}_{n_i}\otimes \mathbf{P}_{\leq i-1})\mathbf{[MY]}^{\langle i \rangle}\mathbf{P}_{\geq i+1}-\mathbf{P}_{\leq i}\mathbf{[MY]}^{\langle i \rangle}\mathbf{P}_{\geq i+1}]+\text{Ten}_d[(\mathbf{I}_{n_d}\otimes \mathbf{P}_{\leq d-1})\mathbf{[MY]}^{\langle d \rangle}],
\end{equation}
where matrices are in bold, with $\mathbf{[MY]}^{\langle i \rangle}\in \mathbb{C}^{n_1...n_i\times n_{i+1}...n_d}$ the $i$-th unfolding of the tensor train $MY\in \mathbb{C}^{n_1\times...\times n_d}$. Here, $\text{Ten}_i$ denotes the construction of the tensor train from its $i$-th unfolded matrix, ${\bf P}_{\leq i-1}$ and ${\bf P}_{\geq i+1}$ denote partial tensor products of tensor cores with indices $\leq i-1$ and $\geq i+1$, respectively (Appendix~\ref{sec:partialproduct}). Appendix~\ref{sec:notation} provides a detailed description of the tensor-train notation, including left- and right-orthogonalization relative to tensor core $i$, which generates $\mathbf{U}_{\leq i}$ and $\mathbf{V}_{\geq i+1}$, with \(\mathbf{P}_{\leq i}=\mathbf{U}_{\leq i}\mathbf{U}_{\leq i}^\dagger\) and \(\mathbf{P}_{\geq i+1}=\mathbf{V}_{\geq i+1}\mathbf{V}_{\geq i+1}^\dagger\) the orthogonal projectors onto \(\mathbf{U}_{\leq i}\), and \(\mathbf{V}_{\geq i+1}\), respectively. 

\subsubsection{Projector-splitting integrator}
The TT-KSL method integrates the equation $\dot{Y}(t)=P_Y(MY(t))$ by splitting $P_Y$. The specific splitting scheme can be introduced with the following notation. Denoting $P_i^+(MY)=\text{Ten}_i[(\mathbf{I}_{n_i}\otimes \mathbf{P}_{\leq i-1})\mathbf{[MY]}^{\langle i \rangle}\mathbf{P}_{\geq i+1}]$, $P_i^-(MY)=\text{Ten}_i[-\mathbf{P}_{\leq i}\mathbf{[MY]}^{\langle i \rangle}\mathbf{P}_{\geq i+1}]$ for $i=1,...,d-1$, and $P_d^+(MY)=\text{Ten}_d[(\mathbf{I}_{n_d}\otimes \mathbf{P}_{\leq d-1})\mathbf{[MY]}^{\langle d \rangle}]$, we rewrite Eq.~\eqref{PXZmain}, as follows:
\begin{equation}\label{simplenotationmain}
\begin{split}
\dot{Y}&=P_Y(MY)\\&=P_1^+(MY)-P_1^-(MY)+P_2^+(MY)-P_2^-(MY)+\cdots -P_{d-1}^-(MY)+P_d^+(MY), 
\end{split}
\end{equation}
where the right-hand side of Eq.~\eqref{simplenotationmain} is a sum of $2d-1$ terms, which can be implemented as the following sequence of initial value problems on the time interval $[t_0,t_1]$: 
\begin{equation}\label{Trottermain}
\begin{split}
&\dot{Y}_1^+=P_1^+(MY_1^+), \; Y_1^+(t_0)=Y_0\\
&\dot{Y}_1^-=P_1^-(MY_1^-), \; Y_1^-(t_0)=Y_1^+(t_1)\\
&\; \; \; \vdots\\
&\dot{Y}_i^+=P_i^+(MY_i^+), \; Y_i^+(t_0)=Y_{i-1}^-(t_1)\\
&\dot{Y}_i^-=P_i^-(MY_i^-), \; Y_i^-(t_0)=Y_i^+(t_1)\\
&\; \; \; \vdots\\
&\dot{Y}_d^+=P_d^+(MY_d^+), \; Y_d^+(t_0)=Y_{d-1}^-(t_1),
\end{split}
\end{equation}
where $t_1=t_0+\tau$ and \(Y_0=Y(t_0)\) is the initial value that corresponds to the wavefunction before the update \(Y(t_1) \approx Y_d^+(t_1)\). The resulting splitting greatly facilitates the integration of $\dot{Y}(t)=P_Y[MY(t)]$ by sequentially updating core-by-core according to $\dot{Y}_i^+$ and $\dot{Y}_i^-$, as shown in Appendix~\ref{sec:ttksl}.

For $i=2,...,d-1$, all $Y_i^+(t_0)$ are left and right orthogonalized(Appendix~\ref{sec:ttksl}.1):
\begin{equation}
Y_i^+(t_0)^{\langle i\rangle}=\mathbf{U}_{\leq i-1}(t_0)\mathbf{K}_i^<(t_0)\mathbf{V}_{\geq i+1}^T(t_0).
\end{equation}

The solution of the equation $\dot{Y}_i^+=P_i^+(\dot{A})$ can then be written, as follows:
\begin{equation}\label{Yi+}
Y_i^+(t_1)^{\langle i\rangle}=\mathbf{U}_{\leq i-1}(t_0)\mathbf{K}_i^<(t_1)\mathbf{V}_{\geq i+1}^T(t_0),
\end{equation}
with
\begin{equation}\label{KSM}
\mathbf{K}_i^<(t_1)=e^{(t_1-t_0)\mathbf{W}_i}\mathbf{K}_i^<(t_0),
\end{equation}
where $\mathbf{W}_i$ is defined, as follows:
\begin{equation}\label{Wi}
\mathbf{W}_i\mathbf{K}_i^<(t)=(\mathbf{I}_{n_i}\otimes \mathbf{U}_{\leq i-1}^\dagger(t_0))(\mathbf{I}_{n_i}\otimes \bar{\mathbf{U}}_{\leq i-1}(t_0))\overline{\mathbf{K}_i^<}(t)\bar{\mathbf{V}}_{\geq i+1}^\dagger(t_0)\mathbf{V}_{\geq i+1}(t_0).
\end{equation}
Here, $\bar{\mathbf{U}}_{\leq i-1}$ represents the left-unfolding matrices of the first $i-1$ cores of $MY_i^+$ and $\bar{\mathbf{V}}_{\geq i+1}$ represents the right-unfolding matrices of the last $d-i$ cores. $\overline{\mathbf{K}_i^<}(t)$ corresponds to the $i$-th core, obtained by operating the $i$-th core of $M$ ({\em i.e.}, $\mathbf{M}_i$) on $\mathbf{K}_i^<(t)$, with proper reshaping (Appendix~\ref{sec:intmatexp}). According to Eq.~\eqref{Yi+} and \eqref{Wi}, only core $i$ is updated, while the other cores are fixed in time. Therefore, the action of $\mathbf{W}_i$ is seen as a single-core effective Hamiltonian that only updates core $i$, which can be compactly written as an operator on a matrix. The operator exponential $e^{(t_1-t_0)\mathbf{W}_i}\mathbf{K}_i^<(t_0)$ is efficiently evaluated in the Krylov subspace, as implemented in the EXPOKIT package\cite{Sidje1998}. It is worth noting that to obtain the left hand side of Eq.~\eqref{Wi}, one would need to carry out the TT matrix-vector multiplication such as $MY_i^+$, and this multiplication could be facilitated when $M$ is a diagonal TT matrix, which effectively converts into an elementwise TT vector-vector multiplication. 

The update of core $i$ is completed by integration of the differential equation of motion $\dot{Y}_i^-(t)=P_i^-(MY_i^-)$, with $Y_i^-(t_0)=Y_i^+(t_1)$. The initial state is orthogonalized at core $i$, as follows:
\begin{equation}\label{S_i}
\begin{split}
Y_i^-(t_0)^{\langle i \rangle}&=\mathbf{U}_{\leq i-1}(t_0)\mathbf{K}_i^<(t_1)\mathbf{V}_{\geq i+1}^T(t_0),\\
&=\mathbf{U}_{\leq i-1}(t_0)\mathbf{U}_i^<(t_1)\mathbf{S}_i(t_0)\mathbf{V}_{\geq i+1}^T(t_0),\\
&=\mathbf{U}_{\leq i}(t_1)\mathbf{S}_i(t_0)\mathbf{V}_{\geq i+1}^T(t_0),
\end{split}
\end{equation}
where $\mathbf{U}_i^<(t_1)\mathbf{S}_i(t_0)=\mathbf{K}_i^<(t_1)$, with $\mathbf{U}_i^<(t_1)$ and $\mathbf{S}_i(t_0)$ obtained by QR decomposition of $\mathbf{K}_i^<(t_1)$. An approach analogous to Eqs.~\eqref{Yi+}-\eqref{Wi} can now be applied to update $\mathbf{S}_i$ (see Appendix~\ref{sec:ttksl}), giving $Y_i^-(t_1)^{\langle i \rangle}=\mathbf{U}_{\leq i}(t_1)\mathbf{S}_i(t_1)\mathbf{V}_{\geq i+1}^\dagger(t_0)$, which is used for the initial state of the next step of update $(i.e., Y_i^-(t_1)^{\langle i\rangle}=Y_{i+1}^+(t_0)^{\langle i\rangle})$ . 

Having updated the first core ($i=1$), the same procedure is then sequentially applied to update all other cores $i=2$--$d$ according to what is called a `forward sweep' update of the tensor, from $Y(t_0)$ to $Y(t_1)$, which is a first-order integrator. Sweeping in reverse ({\em i.e.}, swapping $1,...,d$ with $d,...,1$ in Eq.~\eqref{Trottermain}) results in an alternative first-order integrator, called a `backward sweep'. Combining a forward and a backward sweep with half a time step results in a symmetric, time-reversible second-order integrator.\cite{Lubich2015} That second-order KSL integrator is indicated in Eq.~(\ref{TTSOKSLmain}), as follows:
\begin{equation}\label{KSLroutine}
Y(t_1) = \text{KSL}\left(t_1-t_0, M, Y(t_0)\right), 
\end{equation}
where $Y(t_0)$ is the state to be updated, $\tau=t_1-t_0$ is the time-step, and $M$ is defined by either the kinetic or potential energy term of the Hamiltonian according to the corresponding steps of the SOFT propagation defined in Eqs.~\eqref{eq:SOKSLsplit1}--\eqref{eq:SOKSLsplit3}.

\subsubsection{TT-SOKSL rank-adaptive scheme}
\label{sec:SOKSLRA}
We adapt the rank during each TT-KSL substep of the TT-SOKSL scheme, introduced by Eq.~(\ref{TTSOKSLmain}), to evolve the state with the minimum rank that does not compromise the accuracy. After obtaining $Y(t_1)$ from $Y(t_0)\in \mathcal{M}_r$, 
the propagation is repeated from an initial state with augmented rank $\tilde{Y}(t_0)\in \mathcal{M}_{r+1}$ to obtain $\tilde{Y}(t_1)$, where $\tilde{Y}(t_0)$ is obtained by adding to $Y(t_0)$ a random tensor train of fixed rank (e.g., rank-1) and very small norm (e.g., $10^{-12}$). If the overlap of $\tilde{Y}(t_1)$ and $Y(t_1)$ is sufficiently close to unity, $Y(t_1)$ is used as the time-evolved state. Otherwise, $\tilde{Y}(t_1)$ is used as the initial condition for the next propagation time step, which thus provides rank adaptivity. 

\section{Results}
\begin{figure}[H]
\includegraphics[scale=0.6]{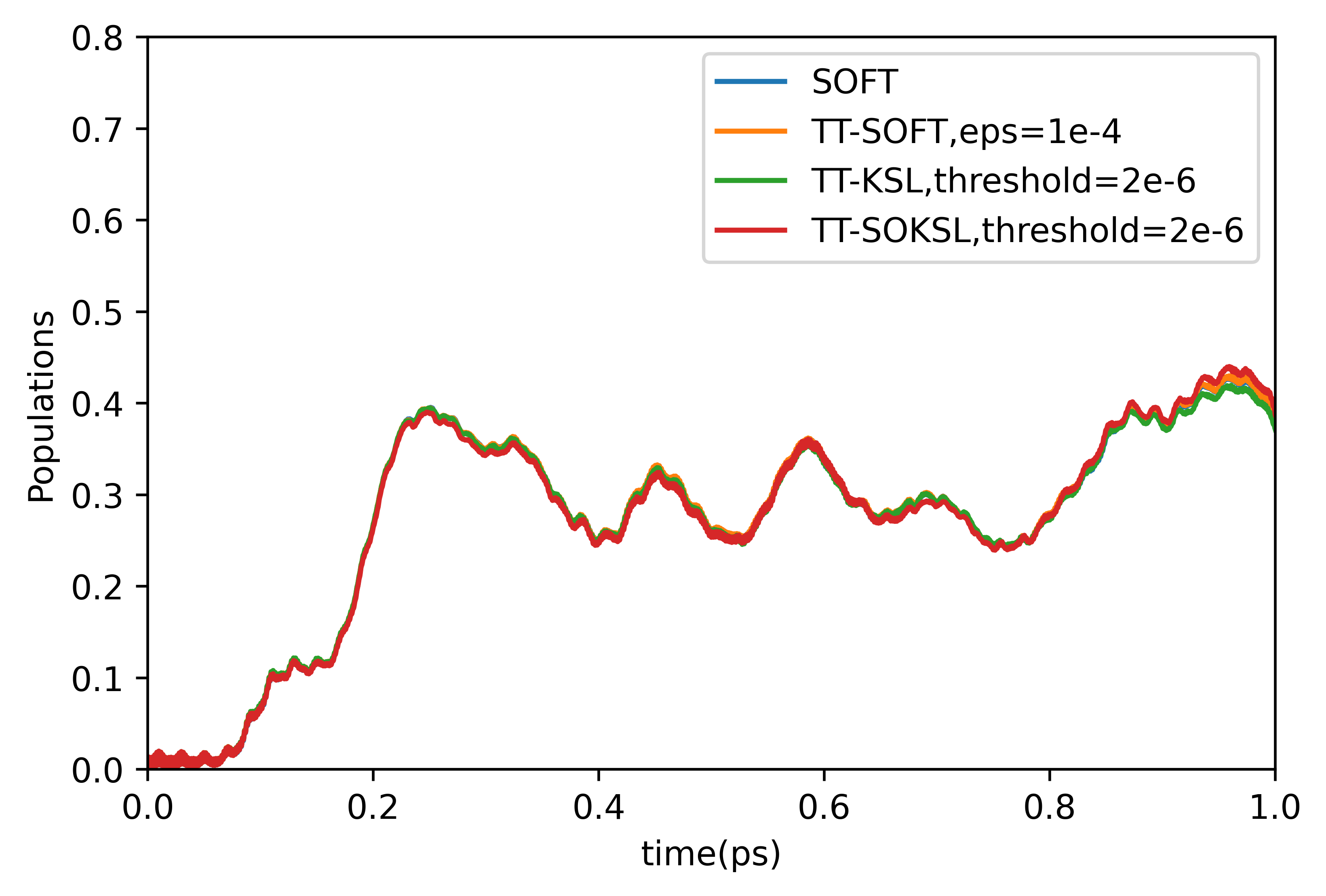}
\includegraphics[scale=0.6]{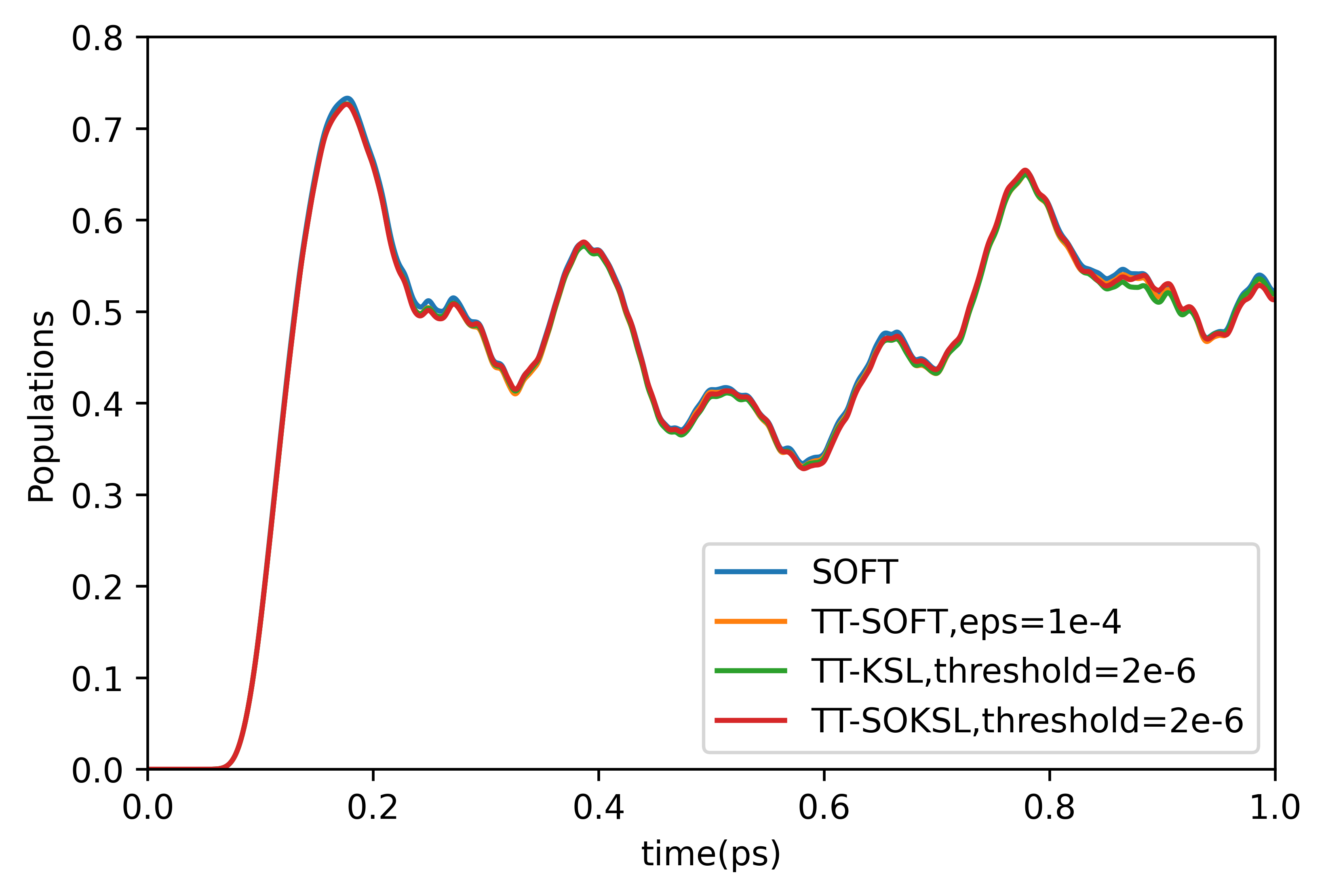}
\includegraphics[scale=0.6]{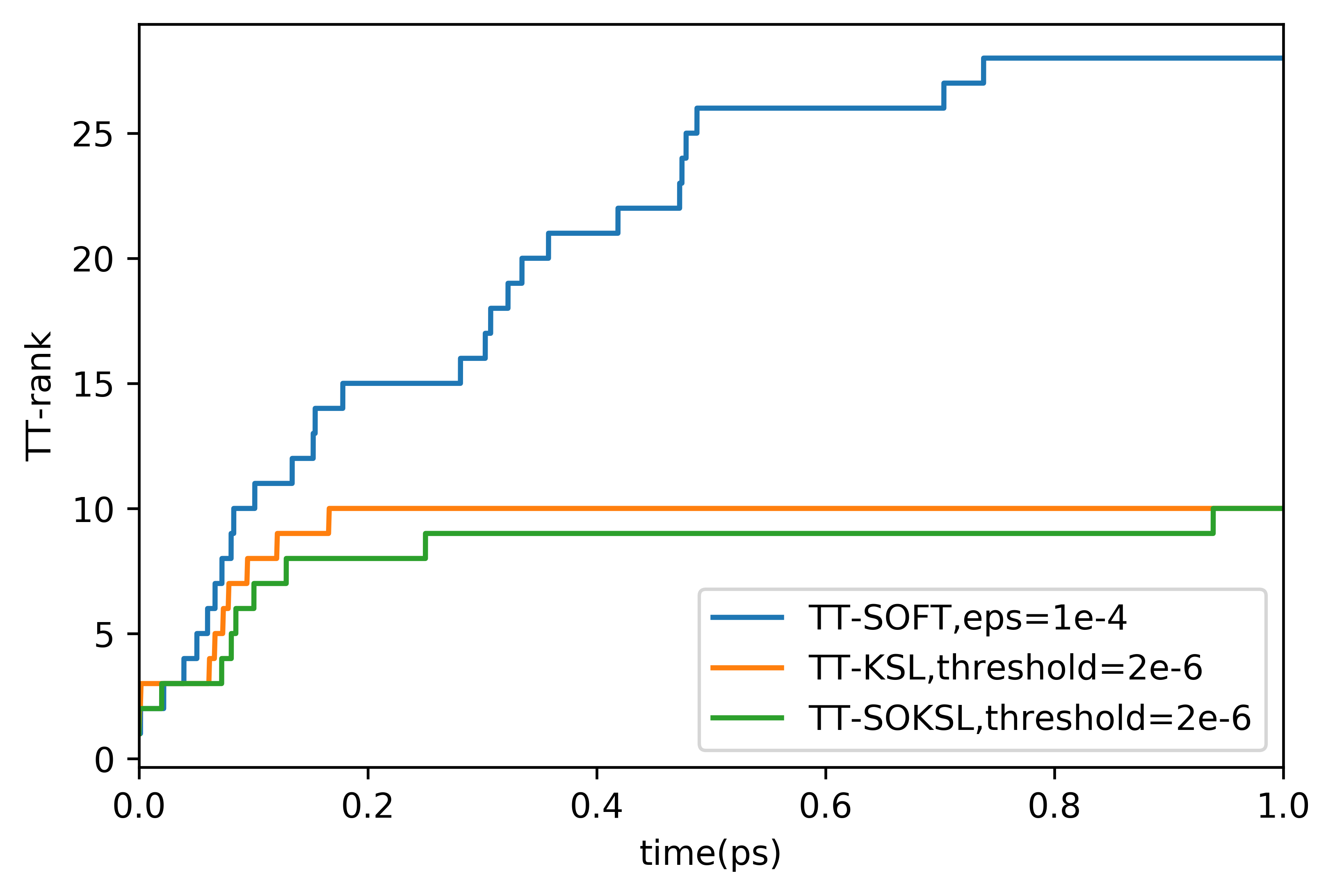}
\caption{(a) Trans population of the 2D retinal model. (b) Diabatic ground state population of the 2D retinal model. (c) TT-rank structure of the 2D retinal simulation. eps is the accuracy parameter of TT rounding, described in Ref.\cite{Greene2017} and threshold$=\langle \tilde{Y}(t_1)|Y(t_1)\rangle-1$ as described in Sec.~\ref{sec:SOKSLRA}. }
\label{fig:2Dpop}
\end{figure}

Figure \ref{fig:2Dpop} (a) and (b) compare the TT calculations to numerically exact full-grid quantum dynamics simulations in simulating the $trans$ $(S_0+S_1)$ and ground state population for a two-dimensional model of the cis/trans isomerization of the retinyl chromophore. 

Figure \ref{fig:2Dpop} shows that transitions to the ground state and formation of the trans isomer begin at $\sim$80 fs after photoexcitation. The trans population curve exhibits a rapid growth reaching its maximum at $\sim$180 fs, where about 70\% of the total population is in the trans form, which successfully describes the primary isomerization event for this ultrafast reaction.\cite{Hahn2000a} That primary event is followed by relaxation to the ground state, as shown by both curves of time-dependent populations that continue to exhibit strong oscillations during the first ps of dynamics, in agreement with previous studies.\cite{flores2004model,Chen2007}

Figure \ref{fig:2Dpop}(c) shows the comparative analysis of the time-dependent TT-rank for TT-SOKSL, TT-KSL, and TT-SOFT. Clearly, TT-SOKSL allows for efficient propagation when compared to TT-KSL and TT-SOFT. The TT ranks of all three methods grow steadily during the primary isomerization event. TT-SOKSL and TT-KSL are very comparable and reach a rank of about 10, whereas TT-SOFT requires a higher rank for comparable precision. These results show that the KSL algorithm is able to evolve the time-dependent state with a lower-rank representation than an algorithm based on the implementation of the time-evolution operator by scaling and squaring followed by rounding, as implemented by TT-SOFT.

\begin{figure}[H]
\includegraphics[scale=0.6]{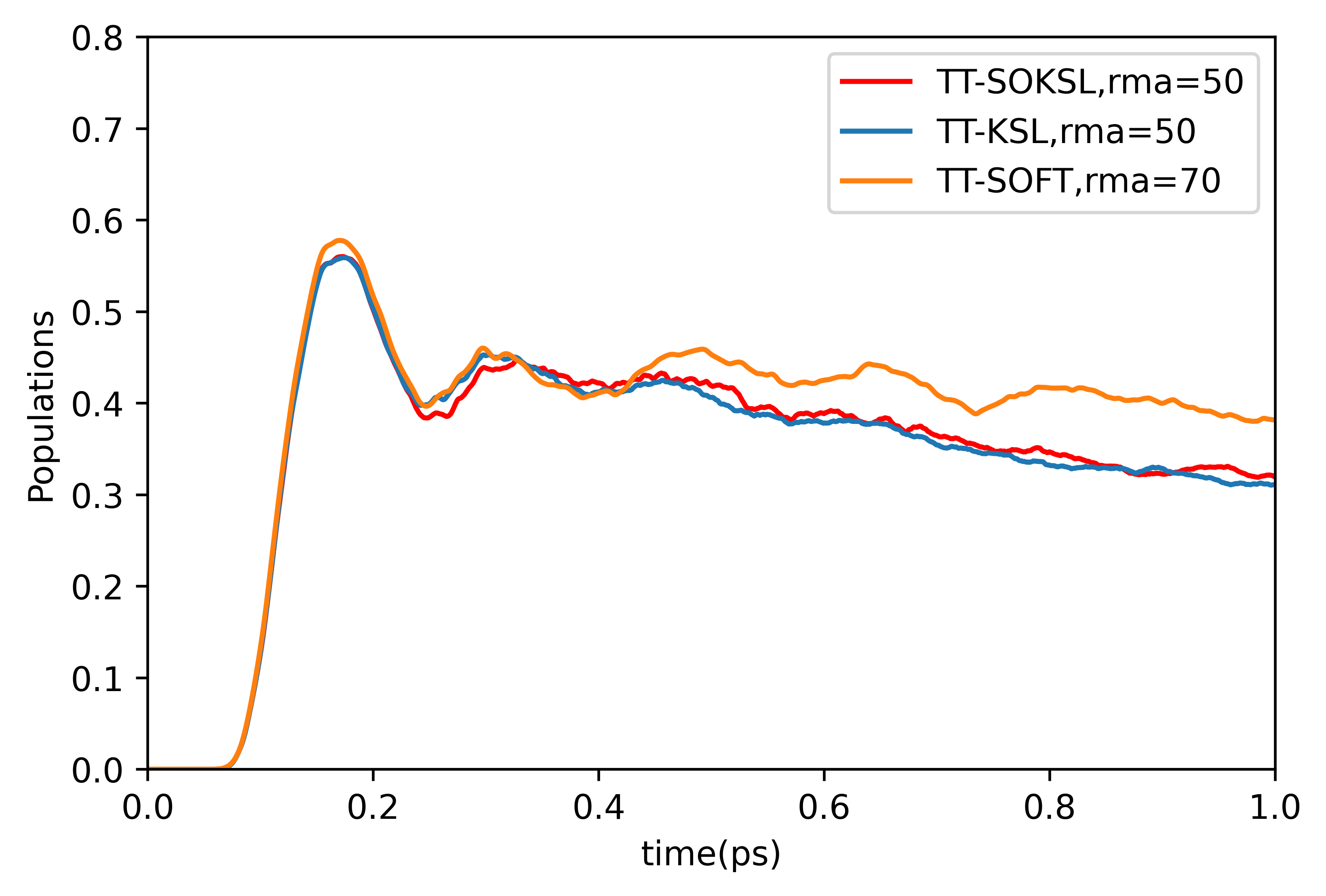}\\
\includegraphics[scale=0.6]{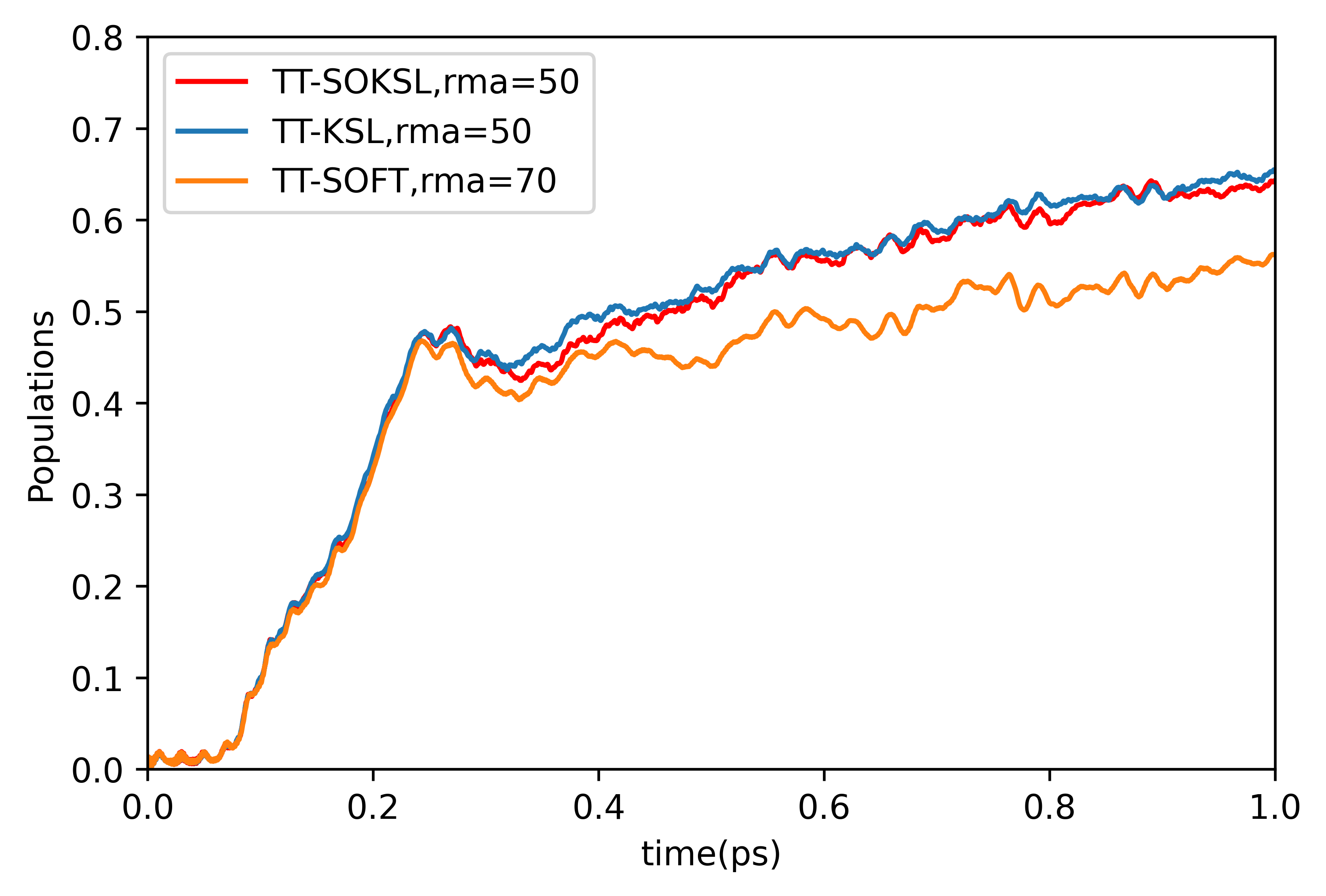}
\caption{(a) Trans population of the 25D retinal model. (b) Diabatic ground state population of the 25D retinal model. rma is the maximally allowed TT-rank, and the reported values are determined by gradual increasing the rank until population curves converge.}
\label{fig:25Dpop}
\end{figure}

Figure \ref{fig:25Dpop} shows the results of simulations for the 25-dimensional (25D) model, showing the capabilities of TT-SOKSL as applied to simulations of non-adiabatic dynamics in high-dimensional model systems.  Figure \ref{fig:25Dpop}(a) shows the time-dependent trans population ($S_0$+$S_1$) and \ref{fig:25Dpop}(b) shows the overall $S_0$ population as they evolve during the first ps of dynamics after photo-excitation to the $S_1$ state. Analysis of the time-dependent populations shows that TT-SOKSL and TT-KSL match closely, whereas TT-SOFT shows some deviations at the longer times. The three methods predict that the main isomerization begins at about $\sim$80 fs after excitation, reaching a maximum of trans population at $\sim$180 fs of dynamics. At 300-1000 fs, after isomerization, all three methods predict a smooth decay of the trans population and a smooth increase of the $S_0$ population, in contrast to the strong oscillations observed in the 2D model. In summary, the analysis of population dynamics shown in Figure \ref{fig:25Dpop}, clearly shows that the results obtained with TT-SOKSL agree very well with those obtained with the state-of-the-art TT-KSL method. 




\begin{figure}[H]
\includegraphics[scale=0.45]{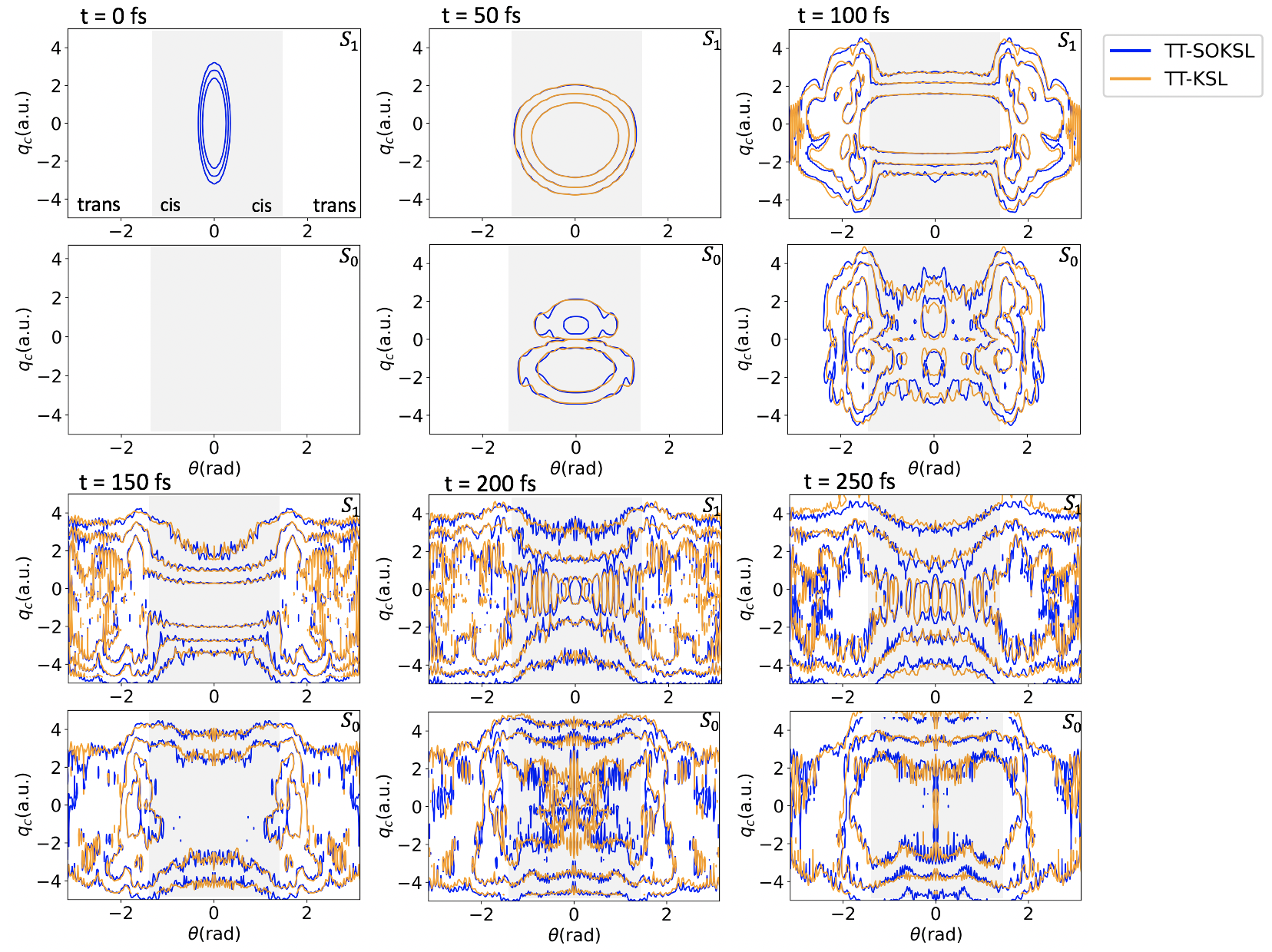}
\caption{Level plot (levels=$10^{-7},10^{-6},10^{-5}$) of the reduced probability density, $\rho(t;\theta, q_c)=\int dz \Psi(t;\theta,q_c,z)^*\Psi(t;\theta,q_c,z)$, over the two primary reactive coordinates, where $z=q_1,...,q_{23}$. The cis regime is shaded.}
\label{fig:2Dwfn}
\end{figure}

Figure \ref{fig:2Dwfn} shows a detailed comparison of the time-dependent reduced probability densities obtained with TT-SOKSL and TT-KSL as a function of the two large amplitude coordinates $\theta$ and $q_c$, after integrating out the bath degrees of freedom. The results show that the wavepacket dynamics is essentially identical for both methods, as shown by the reduced probability density, $\rho(t;\theta, q_c)=\int dz\Psi(t;\theta,q_c,z)^*\Psi(t;\theta,q_c,z)$ in the $S_0$ and $S_1$ electronic states, where $z=q_1,...,q_{23}$ denotes the set of bath coordinates.
Clearly, the probability densities produced by the two methods agree very closely. It is worth noting that, for this particular model system, the $S_1$ wavepacket reaches the edge of the simulation box at 150-200 fs, and strong oscillations arise due to non-adiabatic dynamics. Up to that time, the wavepacket dynamics simulated with TT-SOKSL and TT-KSL agree very well with TT-SOFT, which suggests that small discrepancies at longer times arise due to the truncation scheme of TT-SOFT, which might affect its ability to capture the oscillatory details of the wavepacket. 

\begin{figure*}
\includegraphics[scale=0.6]{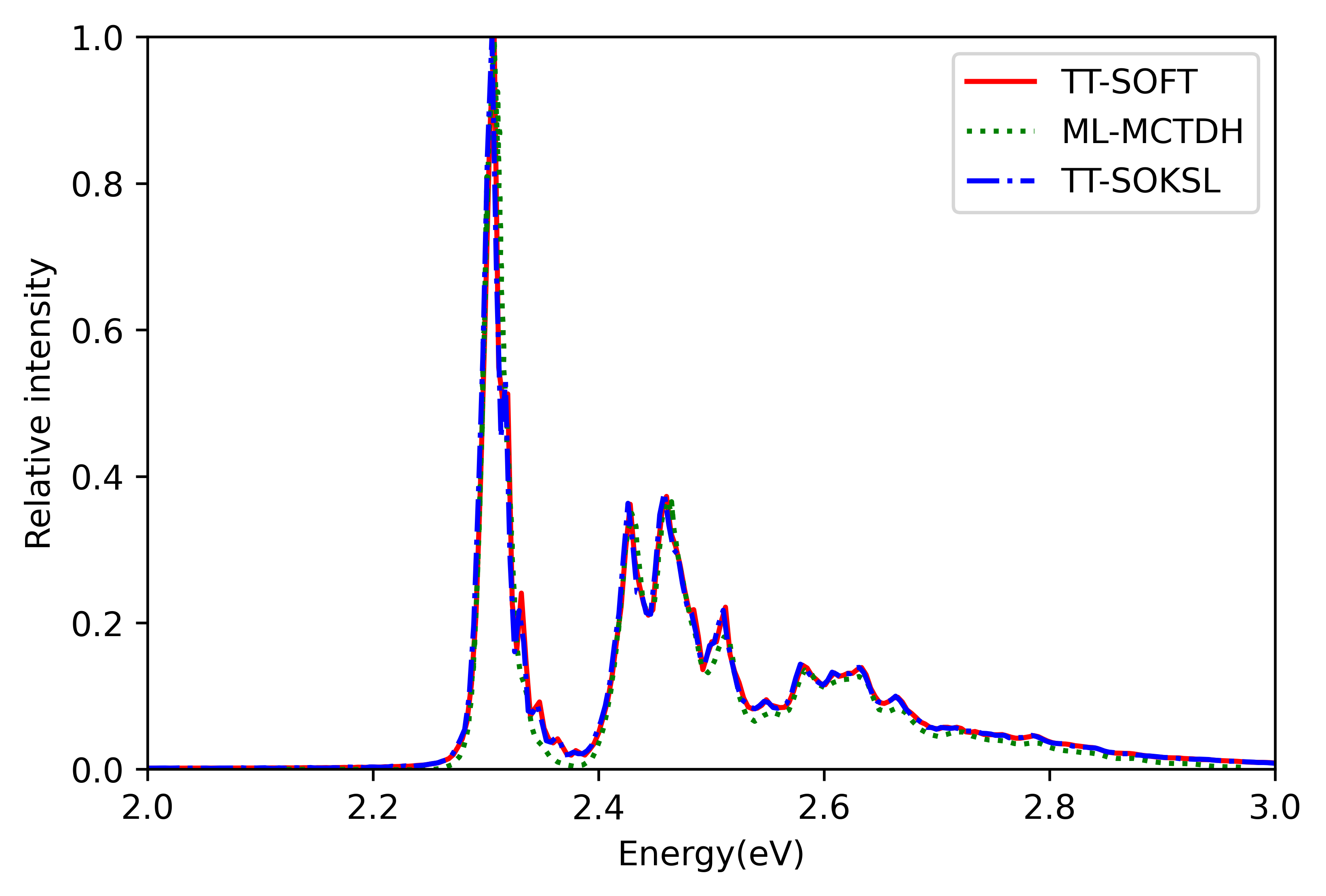}
\caption{Electronic absorption spectrum. ML-MCTDH result reproduced from ref.\cite{Sala2018}.}
\label{fig:25Dspec}
\end{figure*}

Figure \ref{fig:25Dspec} shows the calculated electronic absorption spectrum obtained by TT-SOKSL and TT-SOFT, and ML-MCTDH result\cite{Sala2018}, and shows that the three methods generate a nearly identical electronic absorption spectrum for the 25D retinal model, which further illustrates the capabilities of TT-SOKSL as compared to other state-of-the-art methods. 

\begin{figure*}
\includegraphics[scale=0.6]{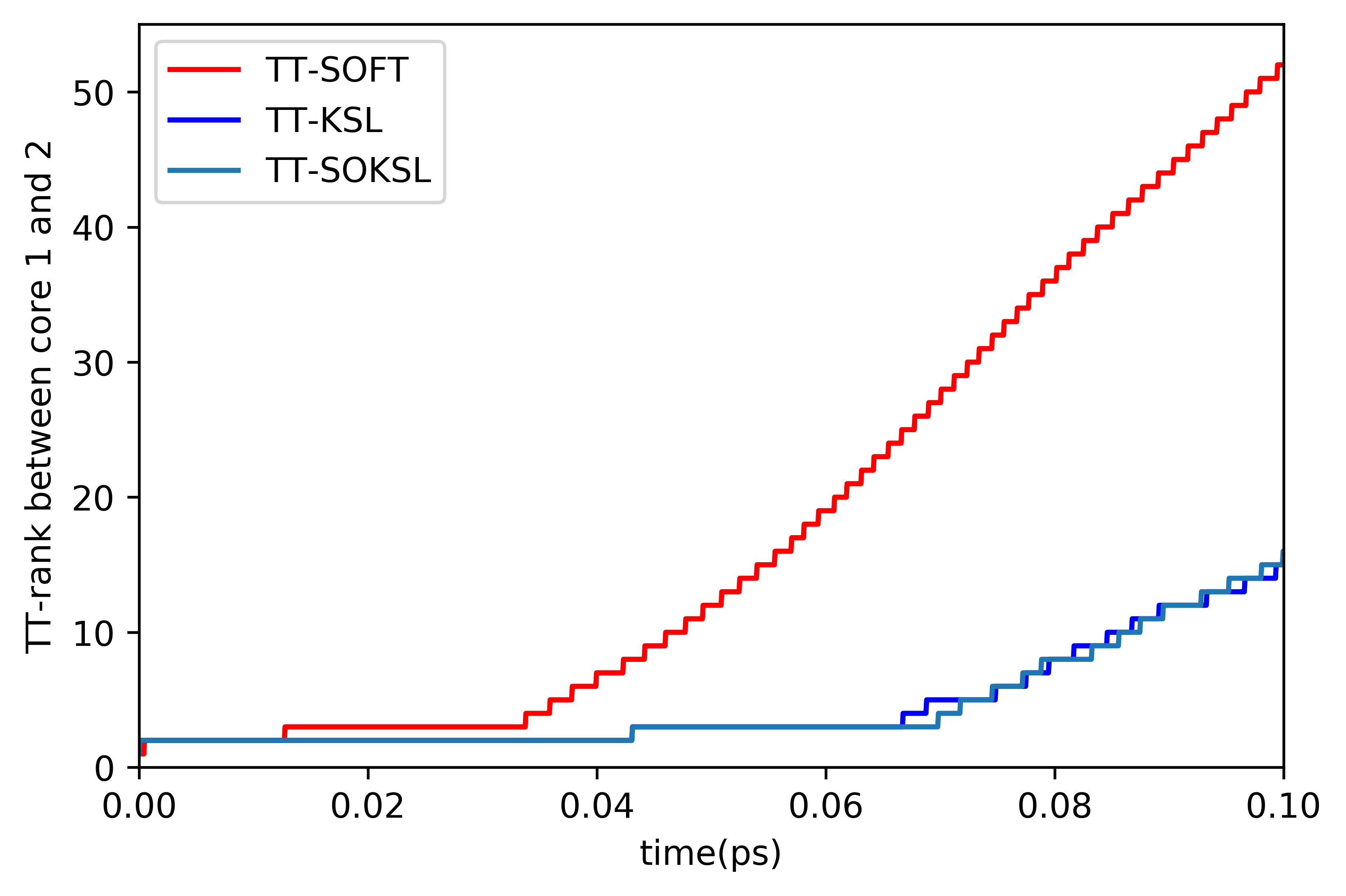}
\caption{TT-rank of the wavefunction for the 25D retinal model at short time.}
\label{fig:25Drank}
\end{figure*}

Figure \ref{fig:25Drank} shows the analysis of the time-dependent TT-ranks for the simulation of the 25-dimensional model system. As for the two-dimensional model, TT-SOKSL enables lower-rank representation than TT-SOFT. During the first 100 fs, the growth of TT-rank is relatively fast for TT-SOFT, reaching 52 at 100 fs, whereas TT-SOKSL requires only a TT-rank of 16, very similar to TT-KSL. 

\begin{figure*}
\includegraphics[scale=0.6]{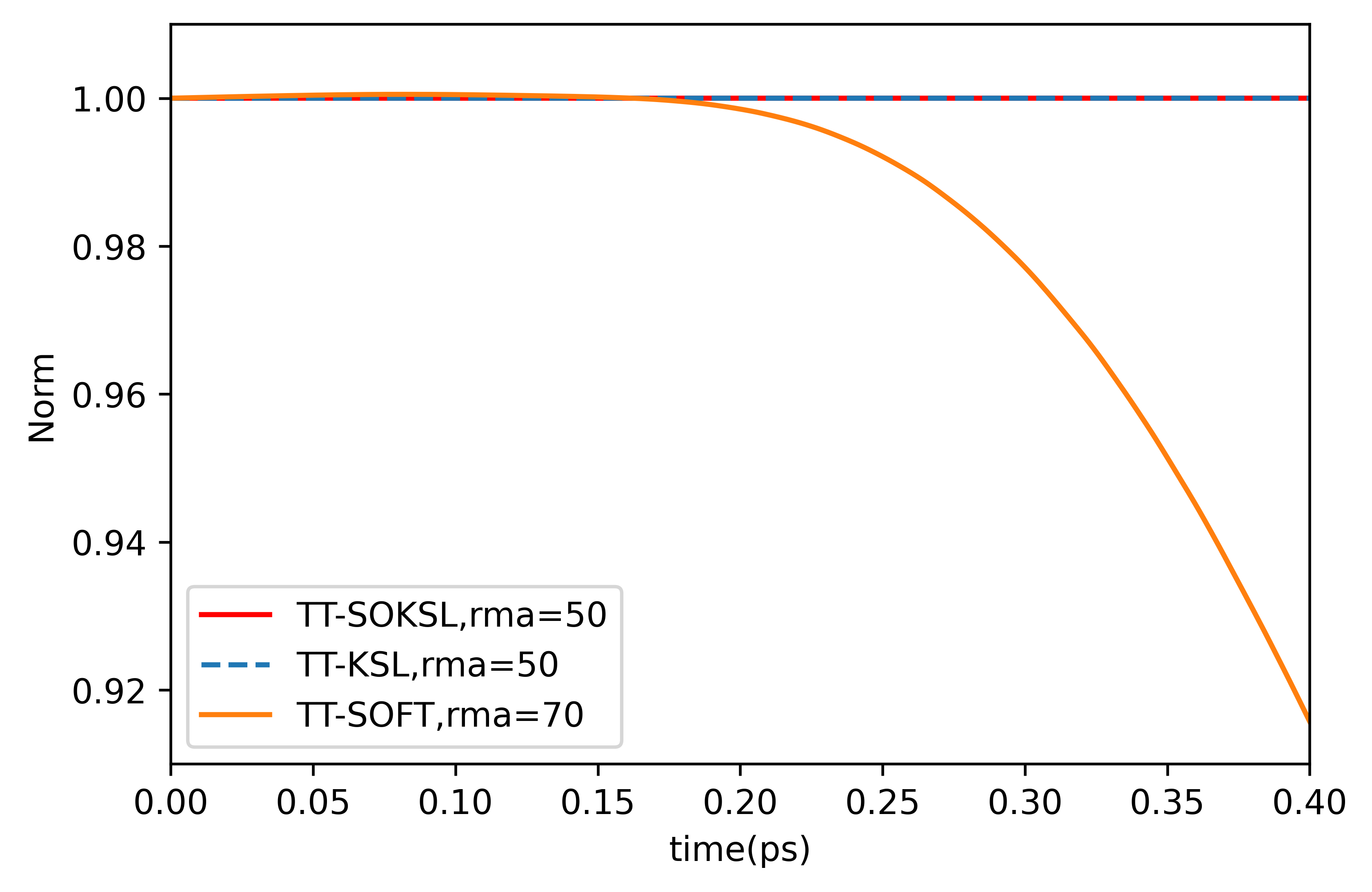}
\caption{Norm conservation for the 25D retinal model.}
\label{fig:25Dnorm}
\end{figure*}

 Figure \ref{fig:25Dnorm} shows the analysis of norm conservation. Clearly, TT-SOKSL agrees well with the level of norm conservation of TT-KSL, whereas TT-SOFT exhibits a significant loss as the propagation proceeds beyond the time when the wavepacket reaches the edge of the simulation box, which suggests superior accuracy of TT-SOKSL for systems with high complexity. 

\section{Discussion and Concluding Remarks}

We have introduced the TT-SOKSL method for quantum simulations of multidimensional model systems. We have demonstrated the capabilities of TT-SOKSL as applied to simulations of non-adiabatic quantum dynamics for a two-state 25-dimensional model system corresponding the photoisomerization of the retinyl chromophore in rhodopsin. TT-SOKSL propagates the quantum state represented as a tensor train using the Trotter expansion of the time-evolution operator as in TT-SOFT.\cite{Greene2017} However, the exponential operators of the Trotter expansion are implemented in TT-SOKSL using the KSL algorithm.\cite{Lubich2015,Lubich2014} Therefore, TT-SOKSL exploits the benefits of both TT-SOFT and TT-KSL methods. Like TT-SOFT, it avoids the need to implement matrix-product operators ({\em e.g.}, the Laplacian operator of the kinetic energy) by transforming the TT representation to momentum space and exploits the linear scaling of multidimensional TT Fourier transforms. Like TT-KSL, it exploits the advantages of projector splitting for efficiently evolving the quantum state on low-rank manifolds. 

We have demonstrated that TT-SOKSL provides two computational advantages over TT-SOFT as applied to simulations of multidimensional quantum dynamics. First, TT-SOKSL allows for more efficient (low-rank) tensor train representations when compared to the propagation scheme based on scaling and squaring followed by rounding in TT-SOFT. Second, TT-SOKSL allows for better norm conservation when applied with limited rank, even when the complexity of the time-dependent state increases ({\em e.g.}, due to non-adiabatic effects and interferences in a periodic potential). Nevertheless, when applied with unlimited maximum rank, both TT-SOKSL and TT-SOFT conserve the norm since they are both formally unitary.\cite{Gatti2017} So, differences in norm conservation when comparing TT-SOKSL, TT-KSL, and TT-SOFT result from the slightly different performance of projection versus rounding to reach a lower-rank manifold.
In TT-SOFT, the wavepacket is rounded after each propagation step according to a maximum TT-rank or desired precision. 
However, the truncation can compromise normalization. On the other hand, the KSL algorithm enforces a fixed rank and the projection onto the tangent space ensures a minimal distance to the true solution. Therefore, no truncation is necessary in TT-KSL and TT-SOKSL and the propagation makes optimal utilization of the low-rank subspace.

We have shown how the TT-SOKSL method avoids the need of a matrix product operator, representing the kinetic energy as a tensor train in momentum space. In contrast to the TT-KSL method that requires a matrix Hamiltonian based on a finite difference or Fourier Grid Hamiltonian such as the DVR, TT-SOKSL operates with diagonal (vector tensor trains) operators. Therefore, TT-SOKSL has the advantage of reduced memory requirement and vector-vector multiplications when compared to the matrix-vector multiplication of TT-KSL. The speed of TT-SOKSL is on par with TT-KSL for the retinal model, as the potential is not diagonal in the electronic degree of freedom and TT-SOKSL propagates three KSL steps (two half potential steps and one kinetic step) for each TT-KSL step. The current implementation requires treatment of the potential matrix as a TT-matrix instead of a TT-vector. We anticipate TT-SOKSL will in fact outperform TT-KSL for single PES problems and for implementations that exploit the sparsity of block-diagonal TT matrices, as encountered in the coupled PES models.

Finally, we note that the strategy of TT-SOKSL --combining the split-operator Hamiltonian and the efficient KSL projection scheme-- could be exploited in other quantum propagation methods. For example, in Chebyshev propagation,\cite{Tal-Ezer1984,Kosloff1993,soley2021functional} the propagator is represented via the Chebyshev expansion, and the Chebyshev polynomials can be represented as a tensor trains.\cite{soley2021functional} So, the TT-SOKSL splitting could provide a more effective scheme to reduce the TT-rank of the Hamiltonian. Solving the split equation with the TT-KSL scheme could provide further computational advantage such as norm conservation and efficient utilization of a low-rank tensor-train array. Therefore, we anticipate the strategy of TT-SOKSL can help facilitate the development of quantum dynamical methods for a wide range of applications. 

\section{Acknowledgements}
V.S.B. acknowledges support from the NSF Grant no. CHE-1900160 and high-performance computing time from NERSC and the Yale High-Performance Computing Center. M.~B.~S.~ acknowledges financial support from the Yale Quantum Institute Postdoctoral Fellowship. N.~L.~ gratefully thanks Professor Xiang Sun (NYU Shanghai) and Professor Haibo Ma (NJU) for stimulating discussions. The authors thank Paul Bergold (TUM) for helpful suggestions. 

\appendix
\section{Dynamical Low-Rank Approximation Method}

The dynamical low-rank approximation method\cite{Koch2007} provides an approximate rank $r$ solution to the differential equation,
\begin{equation}
\dot{\mathbf{\Psi}}(t)=M(\mathbf{\Psi}(t)).
\label{eq:LRA}
\end{equation}
The method can be applied for wavepacket propagation when $\mathbf{\Psi}(t)$ is the time-dependent wavepacket describing the evolution of the system, since Eq.~(\ref{eq:LRA}) is the time-dependent Schr\"odinger equation when $M=-i\hat{H}/\hbar$ is defined in terms of the Hamiltonian $\hat{H}$. In TT-SOKSL, however, the dynamical low-rank approximation is implemented, according to Eq.~(\ref{eq:shr1}), for each of the terms of the Trotter factorization. Therefore, $M$ is defined by either the kinetic, or the potential energy term of the Hamiltonian.

Initially, the wavepacket is reshaped as a
 matrix $\mathbf{Y} \in\mathbb{C}^{m\times n}$ and  decomposed as a rank-$r$ product of three full-rank matrices (e.g., by QR decomposition), as follows: 
\begin{equation}\label{A0yusvv}
\mathbf{Y}(t)=\mathbf{U}(t) \mathbf{S}(t) \mathbf{V}(t)^\dagger,
\end{equation}
where the dagger denotes the adjoint matrix ({\em i.e.}, conjugate transpose), $\mathbf{U}(t) \in \mathbb{C}^{m \times r}$ and $\mathbf{V}(t) \in \mathbb{C}^{n \times r}$ have $r$ orthonormal columns, and $\mathbf{S}(t) \in \mathbb{C}^{r \times r}$ is invertible (not necessarily diagonal).  

We ensure that $\mathbf{Y}(t)$ evolves on the manifold of rank $r$ by propagating the matrices $\mathbf{U}(t)$, $\mathbf{S}(t)$, and $\mathbf{V}(t)$ subject to the orthonormality conditions of Stiefel manifolds, $\mathbf{U}(t)^{\dagger}\mathbf{U}(t)=I$ and $\mathbf{V}(t)^{\dagger} \mathbf{V}(t)=I$. Therefore,
\begin{equation}
\begin{split}
\dot{\mathbf{U}}(t)^{\dagger}\mathbf{U}(t)+c.c&=0,\\
\dot{\mathbf{V}}(t)^{\dagger}\mathbf{V}(t)+c.c.&=0,
\end{split}
\label{eq:stiefelv}
\end{equation}
where c.c. denotes the complex conjugate of the preceding terms. So, the evolution preserves the number $r$ of linearly independent columns of $\mathbf{U}(t)$ and $\mathbf{V}(t)$, which in turn preserves the rank of $\mathbf{Y}(t)$, as shown below with the discussion of Eq.~(\ref{A0tdlravv1}).

There are many ways of enforcing Eq.~(\ref{eq:stiefelv}) and therefore many possible solutions. Nevertheless, a unique solution is obtained by adopting the following gauge conditions:
\begin{equation}
\begin{split}
\dot{\mathbf{U}}(t)^{\dagger}\mathbf{U}(t)&=0,\\
\dot{\mathbf{V}}(t)^{\dagger}\mathbf{V}(t)&=0,
\end{split}
\label{eq:gaugev}
\end{equation}
leading to unique equations of motion for $\mathbf{U}(t)$, $\mathbf{V}(t)$, and $\mathbf{S}(t)$, as shown in Appendix B:
\begin{equation}
\begin{split}
\dot{\mathbf{S}}(t) &= \mathbf{U}^\dagger M(\mathbf{Y}) \mathbf{V},\\
\dot{\mathbf{U}}(t) &=(\mathbf{1}-\mathbf{U}\mathbf{U}^\dagger) M(\mathbf{Y}) \mathbf{V} \mathbf{S}^{-1},\\
\dot{\mathbf{V}}(t) &=(\mathbf{1}-\mathbf{V}\mathbf{V}^\dagger) M(\mathbf{Y})^\dagger \mathbf{U} \mathbf{S}^{-\dagger}.\\
\end{split}
\label{A0dotsv}
\end{equation}

Integrating the equations of motion introduced by Eq.~(\ref{A0dotsv}) by using standard numerical techniques ({\em e.g.}, Runge-Kutta, Velocity-Verlet) becomes challenging when $\mathbf{S}$ has very small singular values since $\mathbf{S}^{-1}$ becomes an ill-conditioned matrix with large norm, so a very small integration time-step is required. In fact, this is what usually happens when the actual rank of the exact solution is smaller than $r$, as in calculations where the rank is not known and thus overestimated to ensure an accurate approximation. A similar problem arises when integrating the equations of motion of the MCTDH method since they depend on the inverse of the density matrix.\cite{Lubich2014a} That problem is usually addressed by some sort of regularization scheme,\cite{regul} although regularization introduces errors that are uncertain. In contrast, the dynamical low-rank approximation method bypasses the need to integrate equations with singular matrices ({\em i.e.}, Eq.~(\ref{A0dotsv})) simply by directly integrating the equation of motion of $\mathbf{Y}$  --{\em i.e.}, Eq.~(\ref{A0tdlravv1}), where there is no singular matrix since $\mathbf{S}^{-1}$ cancels with $\mathbf{S}$. 

The equation of motion for $\mathbf{Y}$ is obtained from Eq.~\eqref{A0yusvv}, as follows:
\begin{equation}\label{A0tdlravv0}
\begin{split}
\dot{\mathbf{Y}}&=\dot{\mathbf{U}} \mathbf{S} \mathbf{V}^\dagger + \mathbf{U} \dot{\mathbf{S}} \mathbf{V}^\dagger + \mathbf{U} \mathbf{S} \dot{\mathbf{V}}^\dagger,
\end{split}
\end{equation}
and, substituting $\dot{\mathbf{U}}$, $\dot{\mathbf{S}}$, and $\dot{\mathbf{V}}$ according to Eq.~\eqref{A0dotsv}, we obtain:
\begin{equation}\label{A0tdlravv10}
\begin{split}
\dot{\mathbf{Y}}
&=(\mathbf{1}-\mathbf{U}\mathbf{U}^\dagger) M(\mathbf{Y}) \mathbf{V} \mathbf{S}^{-1}\mathbf{S} \mathbf{V}^\dagger + \mathbf{U} \mathbf{U}^\dagger M(\mathbf{Y}) \mathbf{V} \mathbf{V}^\dagger + \mathbf{U} \mathbf{S} ((\mathbf{1}-\mathbf{V}\mathbf{V}^\dagger) M(\mathbf{Y})^\dagger \mathbf{U} \mathbf{S}^{-\dagger})^\dagger,
\end{split}
\end{equation}
which gives the equation of motion of $\mathbf{Y}$, as follows:
\begin{equation}\label{A0tdlravv1}
\begin{split}
\dot{\mathbf{Y}}
&=M(\mathbf{Y}) \mathbf{V} \mathbf{V}^\dagger - \mathbf{U} \mathbf{U}^\dagger M(\mathbf{Y})  \mathbf{V} \mathbf{V}^\dagger+ \mathbf{U} \mathbf{U}^\dagger M(\mathbf{Y}),
\end{split}
\end{equation}
Equation~(\ref{A0tdlravv1}) is efficiently integrated by using the Strang splitting approximation, as outlined in Appendix D. Note that each of the terms on the right-hand-side (RHS) of Eq.~(\ref{A0tdlravv1}) involves a projection operator $P_{\mathbf{U}} =\mathbf{U} \mathbf{U}^\dagger$, or $P_{\mathbf{V}}= \mathbf{V} \mathbf{V}^\dagger$, which ensures that $\dot{\mathbf{Y}}$ does not have any component orthogonal to the manifold of rank $r$. The first term corresponds to evolution of $\mathbf{U}$ and $\mathbf{S}$ at constant $\mathbf{V}$. The second term evolves $\mathbf{S}$ at constant $\mathbf{U}$ and $\mathbf{V}$, and the third term  evolves $\mathbf{S}$ and $\mathbf{V}$ at constant $\mathbf{U}$.

\subsection{Projection onto the tangent plane}
Comparing Eq.~\eqref{A0tdlravv1} and Eq.~\eqref{DLRA}, 
we can readily identify $P_\mathbf{Y}$, as follows:
\begin{equation}
\label{eq:py}
\dot{\mathbf{Y}} = P_\mathbf{Y}(M (\mathbf{Y})) = M(\mathbf{Y}) \mathbf{V} \mathbf{V}^\dagger - \mathbf{U} \mathbf{U}^\dagger M(\mathbf{Y})  \mathbf{V} \mathbf{V}^\dagger+ \mathbf{U} \mathbf{U}^\dagger M(\mathbf{Y}).
\end{equation}
Therefore, the projection of any arbitrary state $\mathbf{Z}$ onto the `tangent plane' ({\em i.e.,} the vector space tangent to the manifold of rank $r$ at $\mathbf{Y}$) can be defined, as follows: 
\begin{equation}
\begin{split}
P_\mathbf{Y}(\mathbf{Z})&=\mathbf{Z} \mathbf{V} \mathbf{V}^\dagger-\mathbf{U} \mathbf{U}^\dagger \mathbf{Z} \mathbf{V} \mathbf{V}^\dagger+\mathbf{U} \mathbf{U}^\dagger\mathbf{Z},
\end{split}
\label{eq:proj}
\end{equation}
where the three terms on the RHS of Eq.~\eqref{eq:proj} involve the projection operator $P_{\mathbf{U}}=\mathbf{U}\mathbf{U}^\dagger$, or $P_{\mathbf{V}}=\mathbf{V}\mathbf{V}^\dagger$, so they are on the tangent plane and thus invariant under the effect of $P_\mathbf{Y}$. As an example, we show that $P_\mathbf{Y}$ does not change the middle term of Eq.~\eqref{eq:proj}, as follows:
\begin{equation}
\begin{split}
&P_\mathbf{Y}(-\mathbf{UU}^\dagger \mathbf{Z}\mathbf{VV}^\dagger)\\
=&(-\mathbf{UU}^\dagger \mathbf{Z}\mathbf{VV}^\dagger)\mathbf{VV}^\dagger-\mathbf{UU}^\dagger(-\mathbf{UU}^\dagger \mathbf{Z}\mathbf{VV}^\dagger)\mathbf{VV}^\dagger+\mathbf{UU}^\dagger (-\mathbf{UU}^\dagger \mathbf{Z}\mathbf{VV}^\dagger)\\
=&-\mathbf{UU}^\dagger \mathbf{Z}\mathbf{VV}^\dagger+\mathbf{UU}^\dagger \mathbf{Z}\mathbf{VV}^\dagger-\mathbf{UU}^\dagger \mathbf{Z}\mathbf{VV}^\dagger\\
=&-\mathbf{UU}^\dagger \mathbf{Z}\mathbf{VV}^\dagger.
\end{split}
\end{equation}

Therefore, evolving a rank-$r$ state $\mathbf{Y}$ by displacement along the direction of $\dot{\mathbf{Y}}$, as defined in Eq.~(\ref{A0tdlravv1}), always generates a rank-$r$ state regardless of the rank of $M(\mathbf{Y})$. The resulting propagation avoids the need to first generate a high-rank state $M(\mathbf{Y})$ and then reduce its rank by singular value decomposition or by projection onto the tangent manifold, since $\dot{\mathbf{Y}}$ generates a displacement on the tangent plane.

\section{Equations of motion for \bf{U}, \bf{S} and \bf{V}}

We obtain the equations of motion introduced by Eq.~(\ref{A0dotsv}) by taking the time derivative of Eq.~(\ref{A0yusvv}), as follows:
\begin{equation}
\begin{split}
\dot{\mathbf{Y}}&=
\dot{\mathbf{U}} \mathbf{S} \mathbf{V}^\dagger + \mathbf{U} \dot{\mathbf{S}} \mathbf{V}^\dagger + \mathbf{U} \mathbf{S} \dot{\mathbf{V}}^\dagger,\\
\end{split}
\end{equation}
and imposing the gauge conditions, as follows. 
\subsection{Equation of motion for S}
We invoke orthonormality $\mathbf{V}^\dagger \mathbf{V}=\mathbf{U}^\dagger \mathbf{U}=\mathbf{I}$ with the gauge conditions ({\em i.e.}, $\mathbf{U}^\dagger \dot{\mathbf{U}} = \mathbf{0}$ and $\mathbf{V}^\dagger \dot{\mathbf{V}} = \mathbf{0}$, implying $(\mathbf{V}^\dagger \dot{\mathbf{V}})^\dagger= \dot{\mathbf{V}}^\dagger \mathbf{V}=\mathbf{0}$) to obtain:
\begin{equation}
\begin{split}
\mathbf{U}^\dagger \dot{\mathbf{Y}} \mathbf{V} &=\mathbf{U}^\dagger\dot{\mathbf{U}} \mathbf{S} \mathbf{V}^\dagger \mathbf{V} + \mathbf{U}^\dagger \mathbf{U} \dot{\mathbf{S}} \mathbf{V}^\dagger \mathbf{V} + \mathbf{U}^\dagger \mathbf{U} \mathbf{S} \dot{\mathbf{V}^\dagger}\mathbf{V},\\
&=\mathbf{U}^\dagger\dot{\mathbf{U}} \mathbf{S} \mathbf{V}^\dagger \mathbf{V} + \dot{\mathbf{S}} + \mathbf{U}^\dagger \mathbf{U} \mathbf{S} \dot{\mathbf{V}^\dagger}\mathbf{V}= \dot{\mathbf{S}},\\
\end{split}
\label{eq:sdot}
\end{equation}
Next, we show below that
\begin{equation}
\mathbf{U}^\dagger \dot{\mathbf{Y}} \mathbf{V} = \mathbf{U}^\dagger M(\mathbf{Y}) \mathbf{V},
\label{eq:MY}
\end{equation}
and, substituting Eq.~(\ref{eq:MY}) into Eq.~(\ref{eq:sdot}), we obtain the equation of motion for $\mathbf{S}$, as introduced in Eq.~(\ref{A0dotsv}):
\begin{equation}
\dot{\mathbf{S}}(t) = \mathbf{U}^\dagger M(\mathbf{Y}) \mathbf{V}.
\end{equation}

Equation~(\ref{eq:MY}) is obtained by first noting that the error $\epsilon = M(\mathbf{Y})-\dot{\mathbf{Y}}$ must be orthogonal to any state $\delta \mathbf{Y}$ in the tangent plane --{\em i.e.} $\langle \dot{\mathbf{Y}} - M(\mathbf{Y}) \vert \delta \mathbf{Y} \rangle = 0$. In particular, for the tangent vectors $\delta \mathbf{Y} = \mathbf{U}_i \mathbf{V}_j^\dagger$ with $i, j=1, \dots, r$ (corresponding to vectors of the form $\delta \mathbf{Y} =  {\bf U} \delta{\bf {S}} {\bf V}^{\dagger} + \delta{\bf {U}} {\bf S} {\bf V}^{\dagger} + {\bf U} {\bf S} \delta{\bf {V}}^{\dagger}$, with $\mathbf{U}_i$ the $i^{th}$ column of $\mathbf{U}$, $\mathbf{V}_j$ the $j^{th}$ column of $\mathbf{V}$, $\delta{\bf {U}}=\delta{\bf {V}}=0$, and $\delta{\bf {S}}_{kl}=\delta_{ki} \delta_{lj}$), we obtain:  $\langle \mathbf{U}_i \mathbf{V}_j^\dagger | \dot{\mathbf{Y}}- M(\mathbf{Y}) \rangle = 0$. Rearranging that inner product, we obtain:
\begin{equation}\label{cip}
\begin{split}
\langle \mathbf{U}_i \mathbf{V}_j^\dagger| \dot{\mathbf{Y}} 
\rangle &= \langle \mathbf{U}_i \mathbf{V}_j^\dagger| M(\mathbf{Y}) \rangle.
\end{split}
\end{equation}
Next, we note that $\langle \mathbf{U}_i \mathbf{V}_j^\dagger| {\mathbf{Z}} \rangle=\mathbf{U}_i^\dagger {\mathbf{Z}}\mathbf{V}_j$ where ${\mathbf{Z}}=\{\dot{\mathbf{Y}}, M({\mathbf{Y}})\}$  since
\begin{equation}\label{vecmatvec1}
\begin{split}
\langle \mathbf{U}_i \mathbf{V}_j^\dagger| {\mathbf{Z}} \rangle
&=\sum_{k,l}^{m,n} (\mathbf{U}_i \mathbf{V}_j^\dagger)_{k,l} ^*{\mathbf{Z}}_{k,l}\\
&=\sum_k^m\sum_l^n\mathbf{U}_{ki}^*\mathbf{V}_{jl}{\mathbf{Z}}_{kl}\\
&=\sum_k^m\sum_l^n\mathbf{U}_{ki}^*{\mathbf{Z}}_{kl}\mathbf{V}_{jl}\\
&=\mathbf{U}_{1i}^*\sum_l^n {\mathbf{Z}}_{1l}\mathbf{V}_{jl}+...+\mathbf{U}_{mi}^*\sum_l^n {\mathbf{Z}}_{ml}\mathbf{V}_{jl}\\
&=\mathbf{U}_i^\dagger\begin{pmatrix}
\sum_l^n {\mathbf{Z}}_{1l}\mathbf{V}_{jl}\\
\vdots\\
\sum_l^n {\mathbf{Z}}_{ml}\mathbf{V}_{jl}\\
\end{pmatrix}\\
&=\mathbf{U}_i^\dagger{\mathbf{Z}}\mathbf{V}_j.
\end{split}
\end{equation}
Substituting Eq.~(\ref{vecmatvec1}) into Eq.~(\ref{cip}), we obtain:
\begin{equation}
\mathbf{U}_i^\dagger \dot{\mathbf{Y}} \mathbf{V}_j = \mathbf{U}_i^\dagger M(\mathbf{Y}) \mathbf{V}_j,
\end{equation}
which proves Eq.~(\ref{eq:MY}).


\subsection{Equation of motion for U}
We obtain the equation of motion for ${\mathbf{U}}$, as introduced in Eq.~(\ref{A0dotsv}), by choosing the tangent vector $\delta\mathbf{Y}=\delta u (\sum_j^r\mathbf{S}_{ij}\mathbf{V}_j^\dagger)$, where $\delta u$ is a length-$m$ vector that satisfies the condition $\mathbf{U}^\dagger\delta u=0$, where $\mathbf{V}_j$ is the $j^{th}$ column of $\mathbf{V}$. With this $\delta \mathbf{Y}$, we have the orthogonality condition $\langle M(\mathbf{Y})-\dot{\mathbf{Y}}|\delta u (\sum_j^r\mathbf{S}_{ij}\mathbf{V}_j^\dagger)\rangle=0$, and rearranging the inner product, we obtain:
\begin{equation}\label{dotu1}
\langle\delta u (\sum_j^r\mathbf{S}_{ij}\mathbf{V}_j^\dagger)|\dot{\mathbf{Y}}\rangle=\langle\delta u (\sum_j^r\mathbf{S}_{ij}\mathbf{V}_j^\dagger)|M(\mathbf{Y})\rangle.
\end{equation}
Using the equivalence between the inner product and vector-matrix-vector product, introduced by Eq.~(\ref{vecmatvec1}), the two sides of Eq.~\eqref{dotu1} can be written, as follows:
\begin{equation}\label{dotu2}
\delta u^\dagger\dot{\mathbf{Y}}\sum_j^r\mathbf{S}_{ij}\mathbf{V}_j=\delta u^\dagger M(\mathbf{Y})\sum_j^r\mathbf{S}_{ij}\mathbf{V}_j,
\end{equation}
and introducing the substitution $\sum_j^r\mathbf{S}_{ij}\mathbf{V}_j=\mathbf{V}\mathbf{S}_i$ into Eq.~(\ref{dotu2}), we obtain: 
\begin{equation}
\delta u^\dagger\dot{\mathbf{Y}}\mathbf{V}\mathbf{S}_i=\delta u^\dagger M(\mathbf{Y})\mathbf{V}\mathbf{S}_i.
\end{equation}
Using the condition that $\mathbf{S}$ is invertible, we obtain:
\begin{equation}
\delta u^\dagger\dot{\mathbf{Y}}\mathbf{V}=\delta u^\dagger M(\mathbf{Y})\mathbf{V},
\label{eq:mn0}
\end{equation}
and substituting $\dot{\mathbf{Y}}$ in Eq.~(\ref{eq:mn0}), according to Eq.~\eqref{A0tdlravv0}, we obtain: 
\begin{equation}
\delta u^\dagger(\dot{\mathbf{U}}\mathbf{SV}^\dagger+\mathbf{U}\dot{\mathbf{S}}\mathbf{V}^\dagger+\mathbf{US}\dot{\mathbf{V}}^\dagger)\mathbf{V}=\delta u^\dagger M(\mathbf{Y})\mathbf{V},
\end{equation}
that can be further simplified using $\delta u^\dagger{\mathbf{U}}=0$, as follows:
\begin{equation}
\delta u^\dagger\dot{\mathbf{U}}\mathbf{SV}^\dagger\mathbf{V}=\delta u^\dagger M(\mathbf{Y})\mathbf{V}.
\end{equation}
Further, considering that $\mathbf{V}^\dagger\mathbf{V}=\mathbf{I}$ and $\mathbf{S}$ is invertible, we obtain: 
\begin{equation}\label{dotu4}
\delta u^\dagger\dot{\mathbf{U}}=\delta u^\dagger M(\mathbf{Y})\mathbf{V}\mathbf{S}^{-1},
\end{equation}
so,
\begin{equation}\label{dotu5}
\delta u^\dagger(\dot{\mathbf{U}}-M(\mathbf{Y})\mathbf{V}\mathbf{S}^{-1})=0.
\end{equation}
Considering that Eq.~(\ref{dotu5}) is satisfied by all $\delta u$, so long as $\delta u^\dagger\mathbf{U}=0$, Eq.~(\ref{dotu5}) implies that $\dot{\mathbf{U}}-M(\mathbf{Y})\mathbf{V}\mathbf{S}^{-1}=0$ or  $\dot{\mathbf{U}}-M(\mathbf{Y})\mathbf{V}\mathbf{S}^{-1}=\lambda\mathbf{U}$ where $\lambda$ is a number, so
\begin{equation}\label{dotu6}
(\mathbf{I}-\mathbf{UU}^\dagger)(\dot{\mathbf{U}}-M(\mathbf{Y})\mathbf{VS}^{-1})=0,
\end{equation}
where the effect of the projector $(\mathbf{I}-\mathbf{UU}^\dagger)$ is to enforce the gauge condition  $\delta u^\dagger\mathbf{U}=0$. Rearranging Eq.~\eqref{dotu6}, we obtain:
\begin{equation}\label{dotu7}
(\mathbf{I}-\mathbf{UU}^\dagger)\dot{\mathbf{U}}=(\mathbf{I}-\mathbf{UU}^\dagger)(M(\mathbf{Y})\mathbf{VS}^{-1}),
\end{equation}
and considering that $\mathbf{U}^\dagger\dot{\mathbf{U}}=\mathbf{0}$, we obtain the equation of motion for $\mathbf{U}$, as follows:
\begin{equation}
\dot{\mathbf{U}}=(\mathbf{I}-\mathbf{UU}^\dagger)(M(\mathbf{Y})\mathbf{VS}^{-1}).
\end{equation}
\subsection{Equation of motion for V}
The equation of motion for ${\mathbf{V}}$ is obtained analogously, using the tangent vector $\delta\mathbf{Y}=(\sum_j^r\mathbf{U}_j\mathbf{S}_{ji}) \delta v^\dagger$, with $\delta v$ a length-$n$ vector that satisfies the condition $\mathbf{V}^\dagger\delta v=0$, which must fulfill the orthogonality condition $\langle \delta\mathbf{Y}|\dot{\mathbf{Y}}-M(\mathbf{Y})\rangle=0$:
\begin{equation}
\langle \sum_j^r\mathbf{U}_j\mathbf{S}_{ji} \delta v^\dagger|\dot{\mathbf{Y}}\rangle=\langle \sum_j^r\mathbf{U}_j\mathbf{S}_{ji} \delta v^\dagger|M(\mathbf{Y})\rangle, 
\end{equation}
which can be written as a vector-matrix-vector product, similar to Eq.~\eqref{dotu2}:
\begin{equation}
\label{eq:mn}
(\sum_j^r\mathbf{U}_j^\dagger\mathbf{S}_{ji})\dot{\mathbf{Y}}\delta v=(\sum_j^r\mathbf{U}_j^\dagger\mathbf{S}_{ji})M(\mathbf{Y})\delta v.
\end{equation}
Substituting $\dot{\mathbf{Y}}$ into Eq.~(\ref{eq:mn}), according to Eq.~\eqref{A0tdlravv0}, we obtain: 
\begin{equation}
(\sum_j^r\mathbf{U}_j^\dagger\mathbf{S}_{ji})(\dot{\mathbf{U}}\mathbf{SV}^\dagger+\mathbf{U}\dot{\mathbf{S}}\mathbf{V}^\dagger+\mathbf{US}\dot{\mathbf{V}}^\dagger)\delta v=(\sum_j^r\mathbf{U}_j^\dagger\mathbf{S}_{ji})M(\mathbf{Y})\delta v,
\end{equation}
which is simplified with $\mathbf{V}^\dagger\delta v=0$, as follows:
\begin{equation}
(\sum_j^r\mathbf{U}_j^\dagger\mathbf{S}_{ji})\mathbf{US}\dot{\mathbf{V}}^\dagger\delta v=(\sum_j^r\mathbf{U}_j^\dagger\mathbf{S}_{ji})M(\mathbf{Y})\delta v.
\label{eq:mn2}
\end{equation}
Next, we take the complex transpose (adjoint) of Eq.~(\ref{eq:mn2}) to obtain an explicit expression of $\dot{\mathbf{V}}$, as follows:
\begin{equation}
\delta v^\dagger\dot{\mathbf{V}}\mathbf{S}^\dagger\mathbf{U}^\dagger(\sum_j^r\mathbf{U}_j\mathbf{S}_{ji})=\delta v^\dagger(M(\mathbf{Y}))^\dagger(\sum_j^r\mathbf{U}_j\mathbf{S}_{ji}). 
\end{equation}
and substituting the matrix-vector product, $\sum_j^r\mathbf{U}_j\mathbf{S}_{ji}=\mathbf{US}_i$, we obtain:
\begin{equation}
\delta v^\dagger\dot{\mathbf{V}}\mathbf{S}^\dagger\mathbf{U}^\dagger\mathbf{U}\mathbf{S}_i=\delta v^\dagger(M(\mathbf{Y}))^\dagger\mathbf{U}\mathbf{S}_i, 
\end{equation}
which is further simplified using $\mathbf{U}^\dagger\mathbf{U}=\mathbf{I}$:
\begin{equation}
\delta v^\dagger\dot{\mathbf{V}}\mathbf{S}^\dagger\mathbf{S}_i=\delta v^\dagger(M(\mathbf{Y}))^\dagger\mathbf{U}\mathbf{S}_i. 
\end{equation}
As before, we multiply both sides with $\mathbf{S}_i^{-1}$ to obtain:
\begin{equation}
\delta v^\dagger\dot{\mathbf{V}}\mathbf{S}^\dagger=\delta v^\dagger(M(\mathbf{Y}))^\dagger\mathbf{U}, 
\end{equation}
and multiplying both sides by the inverse of $\mathbf{S}^{\dagger}$ ({\em i.e.}, $\mathbf{S}^{-\dagger}$), we obtain:
\begin{equation}
\delta v^\dagger (\dot{\mathbf{V}}-(M(\mathbf{Y}))^\dagger\mathbf{U}\mathbf{S}^{-\dagger})=0,
\end{equation}
implying that $\dot{\mathbf{V}} - (M(\mathbf{Y}))^\dagger\mathbf{U}\mathbf{S}^{-\dagger}$ is orthogonal to $\delta v^{\dagger}$, or equal to zero. Therefore,
\begin{equation}
(\mathbf{I}-\mathbf{VV}^\dagger) (\dot{\mathbf{V}}-(M(\mathbf{Y}))^\dagger\mathbf{U}\mathbf{S}^{\dagger^{-1}})=0.
\end{equation}
Finally, considering the gauge condition $\mathbf{V}^\dagger\dot{\mathbf{V}}=\mathbf{0}$, we obtain:
\begin{equation}
\dot{\mathbf{V}}=(\mathbf{I}-\mathbf{VV}^\dagger)(M(\mathbf{Y}))^\dagger\mathbf{U}\mathbf{S}^{-\dagger}.
\end{equation}

\subsection{Mutually orthogonal subsets}
The tangent vectors chosen to obtain the equations for ${\mathbf{S}}$, ${\mathbf{U}}$ and ${\mathbf{V}}$ define the subsets,
\begin{equation}\label{mutsub}
\begin{split}
\mathcal{V}_S&=\{\mathbf{U}\delta\mathbf{S}\mathbf{V}^\dagger\},\\
\mathcal{V}_U&=\{\delta \mathbf{USV}^\dagger, \mathbf{U}^\dagger\delta \mathbf{U}=\mathbf{0}\},\\
\mathcal{V}_V&=\{\mathbf{US}\delta\mathbf{V}^\dagger, \mathbf{V}^\dagger\delta \mathbf{V}=\mathbf{0}\},
\end{split}
\end{equation}
which are mutually orthogonal since the inner product between elements from any pair of subsets is equal to zero. For example, $\langle \mathbf{U}\delta\mathbf{S}\mathbf{V}^\dagger \vert \delta \mathbf{USV}^\dagger \rangle=\text{Tr}[ \mathbf{V}\delta\mathbf{S}^\dagger\mathbf{U}^\dagger \delta \mathbf{USV}^\dagger]=0$ since $\mathbf{U}^\dagger\delta\mathbf{U}=0$. Analogously, $\langle \mathbf{U}\delta\mathbf{S}\mathbf{V}^\dagger \vert  \mathbf{US}^\dagger \delta \mathbf{V}^\dagger \rangle=\text{Tr}[ \mathbf{V}\delta\mathbf{S}^\dagger\mathbf{U}^\dagger \mathbf{US}^\dagger \delta \mathbf{V}^\dagger ]=\text{Tr}[\delta \mathbf{V}^\dagger \mathbf{V}\delta\mathbf{S}^\dagger\mathbf{U}^\dagger \mathbf{US}^\dagger  ]=0$ since $\delta\mathbf{V}^\dagger \mathbf{V}=0$. Finally, $\langle \delta\mathbf{U}\mathbf{S}\mathbf{V}^\dagger \vert  \mathbf{US}^\dagger \delta \mathbf{V}^\dagger \rangle=\text{Tr}[ \mathbf{V}\mathbf{S}^\dagger \delta\mathbf{U}^\dagger \mathbf{US}^\dagger \delta \mathbf{V}^\dagger ]=0$ since $\delta\mathbf{U}^\dagger \mathbf{U}=0$.

Together, the three subsets define the complete set of tangent vectors $T_{\mathbf{Y}}\mathcal{M}_r^{m\times n}$, according to the following direct sum of mutually orthogonal subsets:
\begin{equation}\label{defTYMr}
\begin{split}
T_{\mathbf{Y}}\mathcal{M}_r^{m\times n}
&=\mathcal{V}_U\oplus \mathcal{V}_S \oplus \mathcal{V}_V,\\
&=\{\delta \mathbf{Y}=\delta \mathbf{USV}^\dagger+\mathbf{U}\delta \mathbf{S} \mathbf{V}^\dagger+\mathbf{US}\delta \mathbf{V}^\dagger, \mathbf{U}^\dagger \delta \mathbf{U}=\mathbf{0}, \mathbf{V}^\dagger \delta \mathbf{V}=\mathbf{0}\},
\end{split}
\end{equation}
where the curly bracket denotes the set of tangent vectors, with $\delta \mathbf{S}\in \mathbb{C}^{r\times r}$, $\delta\mathbf{U}\in \mathbb{C}^{m\times r}$, and $\delta \mathbf{V}\in\mathbb{C}^{n\times r}$ fulfilling the gauge conditions.

The mutually orthogonal relationship unifies the derivation of the equations of motion for ${\mathbf{U}}$, ${\mathbf{S}}$, and ${\mathbf{V}}$, as follows: 
\begin{equation}
\label{eq:ortho}
\langle M(\mathbf{Y})-\dot{\mathbf{Y}}| \delta \mathbf{Y} \rangle = 0\;\;\text{for any}\;\delta \mathbf{Y}\in T_{\mathbf{Y}}\mathcal{M}_r. 
\end{equation}
When $\delta \mathbf{Y}=\delta \mathbf{USV}^\dagger$, \begin{equation}\label{TDVPu}
\begin{split}
\langle M(\mathbf{Y})-\dot{\mathbf{Y}} | \delta \mathbf{USV}^\dagger\rangle &=\langle M(\mathbf{Y})-\dot{\mathbf{U}} \mathbf{S} \mathbf{V}^\dagger - \mathbf{U} \dot{\mathbf{S}} \mathbf{V}^\dagger - \mathbf{U} \mathbf{S} \dot{\mathbf{V}}^\dagger  | \delta \mathbf{USV}^\dagger\rangle=0,\\
\end{split}
\end{equation}
and considering that $\mathbf{U}^\dagger\delta\mathbf{U}=\mathbf{0}$, 
$\langle \mathbf{U} \dot{\mathbf{S}} \mathbf{V}^\dagger - \mathbf{U} \mathbf{S} \dot{\mathbf{V}}^\dagger  | \delta \mathbf{USV}^\dagger\rangle = \langle  \mathbf{U} ( \dot{\mathbf{S}} \mathbf{V}^\dagger - \mathbf{S} \dot{\mathbf{V}}^\dagger)  | \delta \mathbf{USV}^\dagger\rangle = \text{Tr}[ ( \dot{\mathbf{V}} \mathbf{S}^\dagger - \mathbf{V} \dot{\mathbf{S}}^\dagger) \mathbf{U}^\dagger \delta \mathbf{USV}^\dagger] = 0$, we obtain:
\begin{equation}\label{TDVPu2}
\begin{split}
&\langle M(\mathbf{Y})-\dot{\mathbf{Y}} | \delta \mathbf{USV}^\dagger\rangle = \langle M(\mathbf{Y})-\dot{\mathbf{U}} \mathbf{S} \mathbf{V}^\dagger | \delta \mathbf{USV}^\dagger \rangle = 0.
\end{split}
\end{equation}
Rearranging Eq.~\eqref{TDVPu2}, we obtain:
\begin{equation}\label{deridelu}
\begin{split}
\langle M(\mathbf{Y})-\dot{\mathbf{U}} \mathbf{S} \mathbf{V}^\dagger | \delta \mathbf{USV}^\dagger \rangle
=&\langle M(\mathbf{Y}) | \delta \mathbf{USV}^\dagger \rangle-\langle \dot{\mathbf{U}} \mathbf{S} \mathbf{V}^\dagger | \delta \mathbf{USV}^\dagger \rangle\\
=&\text{Tr}[M(\mathbf{Y})^\dagger  \delta \mathbf{USV}^\dagger]-\text{Tr}[\mathbf{V}\mathbf{S}^\dagger\dot{\mathbf{U}}^\dagger\delta \mathbf{USV}^\dagger]\\
=&\text{Tr}[\mathbf{SV}^\dagger M(\mathbf{Y})^\dagger  \delta \mathbf{U}]-\text{Tr}[\mathbf{SV}^\dagger\mathbf{V}\mathbf{S}^\dagger\dot{\mathbf{U}}^\dagger\delta \mathbf{U}]\\
=&\text{Tr}[\mathbf{SV}^\dagger M(\mathbf{Y})^\dagger  \delta \mathbf{U}]-\text{Tr}[\mathbf{S}\mathbf{S}^\dagger\dot{\mathbf{U}}^\dagger\delta \mathbf{U}]\\
=&\langle M(\mathbf{Y})\mathbf{V}\mathbf{S}^\dagger|\delta \mathbf{U}\rangle-\langle \dot{\mathbf{U}}\mathbf{S}\mathbf{S}^\dagger|\delta \mathbf{U}\rangle\\
=&\langle M(\mathbf{Y})\mathbf{V}\mathbf{S}^\dagger-\dot{\mathbf{U}}\mathbf{S}\mathbf{S}^\dagger|\delta \mathbf{U}\rangle,
\end{split}
\end{equation}
so, according to Eqs.~(\ref{TDVPu2}) and~(\ref{deridelu}), $\langle M(\mathbf{Y})\mathbf{V}\mathbf{S}^\dagger-\dot{\mathbf{U}}\mathbf{S}\mathbf{S}^\dagger|\delta \mathbf{U}\rangle=0$, which implies that $M(\mathbf{Y})\mathbf{V}\mathbf{S}^\dagger-\dot{\mathbf{U}}\mathbf{S}\mathbf{S}^\dagger$ is orthogonal to $\delta \mathbf{U}$ ({\em i.e.}, parallel to $\mathbf{U}$ since $\mathbf{U}^\dagger \delta \mathbf{U}=\mathbf{0}$). So,
\begin{equation}\label{projdotu}
\begin{split}
(\mathbf{I}-\mathbf{UU}^\dagger)  (M(\mathbf{Y})\mathbf{V}\mathbf{S}^\dagger-\dot{\mathbf{U}}\mathbf{S}\mathbf{S}^\dagger)
&=(\mathbf{I}-\mathbf{UU}^\dagger)  M(\mathbf{Y})\mathbf{V}\mathbf{S}^\dagger-\dot{\mathbf{U}}\mathbf{S}\mathbf{S}^\dagger=0,
\end{split}
\end{equation}
where we used the gauge condition $\mathbf{U}^\dagger\dot{\mathbf{U}}=\mathbf{0}$. Rearranging the second equality of Eq.~\eqref{projdotu} leads to the equation of motion for $\mathbf{U}$:
\begin{equation}
\begin{split}
&\dot{\mathbf{U}}\mathbf{S}\mathbf{S}^\dagger=(\mathbf{I}-\mathbf{U} \mathbf{U}^\dagger) M(\mathbf{Y}) \mathbf{V} \mathbf{S}^\dagger,\\
&\dot{\mathbf{U}}=(\mathbf{I}-\mathbf{U} \mathbf{U}^\dagger) M(\mathbf{Y}) \mathbf{V} \mathbf{S}^{-1}. 
\end{split}
\end{equation}
Analogously, choosing $\delta \mathbf{Y}=\mathbf{U}\delta \mathbf{S}\mathbf{V}^\dagger$ in Eq.~\eqref{eq:ortho} leads to: 
\begin{equation}
\langle M(\mathbf{Y})-\mathbf{U} \dot{\mathbf{S}} \mathbf{V}^\dagger | \mathbf{U}\delta \mathbf{S}\mathbf{V}^\dagger \rangle = 0,
\end{equation}
which, with a procedure similar to Eq.~\eqref{deridelu}, leads to:
\begin{equation}\label{deridels}
\langle\dot{\mathbf{S}} |\delta \mathbf{S}\rangle=\langle \mathbf{U}^\dagger M(\mathbf{Y})\mathbf{V}|\delta \mathbf{S}\rangle.
\end{equation}
Therefore,
\begin{equation}
\dot{\mathbf{S}}=\mathbf{U}^\dagger M(\mathbf{Y})\mathbf{V}. 
\end{equation}
As for the derivation of $\dot{\mathbf{U}}$, choosing $\delta \mathbf{Y}=\mathbf{U}\mathbf{S}\delta\mathbf{V}^\dagger$ in Eq.~\eqref{eq:ortho} gives the equation for $\mathbf{V}$: 
\begin{equation}
\dot{\mathbf{V}}=(\mathbf{I}-\mathbf{V} \mathbf{V}^\dagger) M(\mathbf{Y})^\dagger \mathbf{U} \mathbf{S}^{-\dagger}.
\end{equation}

\section{KSL Integration}
\label{sec:kslintegration}
The equation of motion for ${\mathbf{Y}}$, introduced by Eq.~(\ref{eq:py}),
\begin{equation}
\begin{split}
\dot{\mathbf{Y}} 
&= M(\mathbf{Y}) \mathbf{V}\mathbf{V}^\dagger - \mathbf{U}\mathbf{U}^\dagger M(\mathbf{Y})  \mathbf{V}\mathbf{V}^\dagger+ \mathbf{U}\mathbf{U}^\dagger M( \mathbf{Y}),
\end{split}
\label{eq:sdoot}
\end{equation}
is integrated by sequentially applying the three terms on the RHS of Eq.~(\ref{eq:sdoot}) according to the following Lie-Trotter splitting method over the time interval $[t_0,t_1]$:
\begin{equation}\label{LT}
\begin{split}
&\dot{\mathbf{Y}}_I=M(\mathbf{Y}_0) \mathbf{V}_I\mathbf{V}_I^\dagger,\; \;\;\;\;\;\;\;\;\;\;\;\;\;\;\;\;\;\;\;\;\;\;\text{with~}\mathbf{Y}_I(t_0)=\mathbf{U}_I\mathbf{S}_I\mathbf{V}_I^\dagger=\mathbf{Y}_0\\
&\dot{\mathbf{Y}}_{II}=-\mathbf{U}_{II}\mathbf{U}_{II}^\dagger M( \mathbf{Y}_0)\mathbf{V}_{II}\mathbf{V}_{II}^\dagger,\;\;\;\;\text{with~}\mathbf{Y}_{II}(t_0)=\mathbf{U}_{II}\mathbf{S}_{II}\mathbf{V}_{II}^\dagger=\mathbf{Y}_I(t_1)  \\
&\dot{\mathbf{Y}}_{III}=\mathbf{U}_{III}\mathbf{U}_{III}^\dagger M( \mathbf{Y}_0),\;\;\;\; \;\;\;\;\;\;\;\;\;\;\;\text{with~}\mathbf{Y}_{III}(t_0)=\mathbf{U}_{III}\mathbf{S}_{III}\mathbf{V}_{III}^\dagger(t_0)=\mathbf{Y}_{II}(t_1)
\end{split}
\end{equation}
which gives the approximation $\mathbf{Y}(t_1)\approx \mathbf{Y}_{III}(t_1)$. 
Note that the first equation of motion, introduced by the splitting scheme of Eq.~(\ref{LT}), evolves $\mathbf{Y}=\mathbf{U}\mathbf{S}\mathbf{V}^\dagger$ at constant ${\mathbf{V}}$ ({\em i.e.}, $\dot{\mathbf{V}}_I=0$) due to the effect of projection operator $P_{\mathbf{V}_I}= \mathbf{V}_I \mathbf{V}_I^\dagger$, which effectively propagates only the product ${\mathbf{K}}={\mathbf{U}}{\mathbf{S}}$. Considering that
$\dot{\mathbf{Y}}_I=M(\mathbf{Y}_0) \mathbf{V}_I\mathbf{V}_I^\dagger=\dot{\mathbf{K}}\mathbf{V}_I^\dagger$, we obtain $\dot{\mathbf{K}}=M(\mathbf{Y}_0)\mathbf{V}_I$. The second equation of Eq.~(\ref{LT}) evolves ${\mathbf{S}}$ at constant ${\mathbf{U}}$ and ${\mathbf{V}}$, and the third equation evolves the product ${\mathbf{L}}={\mathbf{S}}{\mathbf{V}^{\dagger}}$ at constant ${\mathbf{U}}$. Therefore, the resulting method avoids equations of motion with singular matrices by sequentially propagating the matrices ${\mathbf{K}}$, ${\mathbf{S}}$, and ${\mathbf{L}}$ ({\em i.e.}, by KSL propagation).


\subsection{Integration of K}
\label{sec:kintegration}
We solve the first equation of motion, introduced by Eq.~\eqref{LT},
\begin{equation}
\dot{\mathbf{Y}}_I = M(\mathbf{Y}_0) \mathbf{V}_I \mathbf{V}_I^\dagger,
\label{eq:st11}
\end{equation}
using the initial condition, $\mathbf{Y}_I(t_0)=\mathbf{Y}_0$ with $\mathbf{Y}_I=\mathbf{U}_I\mathbf{S}_I\mathbf{V}_I^\dagger$, and $\dot{\mathbf{V}}_I=0$, so 
\begin{equation}
\begin{split}
\dot{\mathbf{Y}}_I &=\dot{\mathbf{K}} \mathbf{V}_I^\dagger,
\end{split}
\label{eq:st112}
\end{equation}
with $\mathbf{K}=\mathbf{U}_I\mathbf{S}_I$. Integrating Eq.~(\ref{eq:st112}) with the trapezoidal rule, we obtain:
\begin{equation}\label{int11}
\begin{split}
\int_{t_0}^{t_1}dt \dot{\mathbf{Y}}_I &=\int_{t_0}^{t_1} dt \dot{\mathbf{K}} \mathbf{V}_I^\dagger,\\
{\mathbf{Y}}_I(t_1)-{\mathbf{Y}}_I(t_0) &\approx (t_1-t_0) M(\mathbf{Y}_0) \mathbf{V}_I(t_0) \mathbf{V}_I^\dagger(t_0).
\end{split}
\end{equation}
Therefore, 
\begin{equation}
\mathbf{Y}_I(t_1)\approx(\mathbf{U}_I(t_0)\mathbf{S}_I(t_0)+(t_1-t_0)M(\mathbf{Y}_0)\mathbf{V}_I(t_0))\mathbf{V}^\dagger(t_0).
\end{equation}

\subsection{Integration of S}
\label{sec:sintegration}
Having obtained $\mathbf{Y}_I(t_1)=\mathbf{U}_I(t_1)\mathbf{S}_I(t_1)\mathbf{V}_I(t_0)^\dagger$, we compute $\mathbf{U}_I(t_1)$ and $\mathbf{S}_I(t_1)$  by QR decomposition of $\mathbf{K}_1(t_1)=\mathbf{Y}_I(t_1) \mathbf{V}_I(t_0)$ ({\em i.e.}, $\text{QR}[\mathbf{K}_I(t_1)]\rightarrow \mathbf{U}_I(t_1),\;\mathbf{S}_I(t_1)$), and we initialize $\mathbf{Y}_{II}=\mathbf{U}_{II}\mathbf{S}_{II}\mathbf{V}_{II}^\dagger$ with
$\mathbf{U}_{II}(t_0)=\mathbf{U}_I(t_1)$, $\mathbf{S}_{II}(t_0)=\mathbf{S}_I(t_1)$ and $\mathbf{V}_{II}(t_0)=\mathbf{V}_I(t_0)$. We integrate the equation:
\begin{equation}
\dot{\mathbf{Y}}_{II} = - \mathbf{U}_{II}\mathbf{U}_{II}^\dagger  M(\mathbf{Y}_0) \mathbf{V}_{II}^\dagger,
\label{eq:st2}
\end{equation}
keeping constant $\mathbf{U}_{II}$ and $\mathbf{V}_{II}$, and effectively integrate the equation $\dot{\mathbf{S}}_{II}=- \mathbf{U}_{II}^\dagger M(\mathbf{Y}_0)  \mathbf{V}_{II}$ from $t_0$ to $t_1$ with the trapezoidal rule, as follows:
\begin{equation}
\begin{split}
\mathbf{S}_{II}(t_1)
&\approx \mathbf{S}_{II}(t_0)-\mathbf{U}_{II}^\dagger(t_0)(t_1-t_0)M(\mathbf{Y}_0)\mathbf{V}_{II}(t_0),
\end{split}
\end{equation}
to obtain $\mathbf{Y}_{II}(t_1)$, as follows:
\begin{equation}\label{eq:st22}
\begin{split}
\mathbf{Y}_{II}(t_1)&=\mathbf{U}_{II}(t_0)\mathbf{S}_{II}(t_1)\mathbf{V}_{II}^\dagger(t_0).\\
\end{split}
\end{equation}

\subsection{Integration of L}
\label{sec:lintegration}
Finally, we initialize ${\mathbf{Y}}_{III}=\mathbf{U}_{III}\mathbf{S}_{III}\mathbf{V}_{III}^\dagger$ with $\mathbf{U}_{III}(t_0)=\mathbf{U}_{II}(t_0)$,  $\mathbf{S}_{III}(t_0)=\mathbf{S}_{II}(t_1)$ and  $\mathbf{V}_{III}(t_0)=\mathbf{V}_{II}(t_0)$, and we solve the equation
$\dot{\mathbf{Y}}_{III} = \mathbf{U}_{III}\mathbf{U}_{III}^\dagger {M(\mathbf{Y}_0)}$,
with constant $\mathbf{U}_{III}$, which effectively propagates only the product $\mathbf{L}=\mathbf{S}_{III}\mathbf{V}_{III}^\dagger$. Considering that 
\begin{equation}
\label{eq:yiii}
\dot{\mathbf{Y}}_{III} = \mathbf{U}_{III} \dot{\mathbf{L}},
\end{equation}
we obtain $\dot{\mathbf{L}}=\mathbf{U}_{III}^\dagger {M(\mathbf{Y}_0)}$, which upon integration by the trapezoidal rule gives
\begin{equation}
\begin{split}
{\mathbf{L}}(t_1) &\approx {\mathbf{L}}(t_0) + \mathbf{U}_{III}(t_0)^\dagger M(\mathbf{Y}_0) (t_1-t_0),\\
&=\mathbf{S}_{III}(t_1) \mathbf{V}_{III}(t_1)^{\dagger},
\end{split}
\end{equation}
with $\mathbf{S}_{III}(t_1)$ and $\mathbf{V}_{III}(t_1)$ obtained by QR decomposition of ${\mathbf{L}}(t_1)$.

Integrating Eq.~(\ref{eq:yiii}), we obtain:
\begin{equation}
\begin{split}
{\mathbf{Y}}_{III}(t_1) - {\mathbf{Y}}_{III}(t_0) = \mathbf{U}_{III}(t_0) ({\mathbf{L}}(t_1)-{\mathbf{L}}(t_0)),\\
=\mathbf{U}_{III}(t_0) \mathbf{U}_{III}(t_0)^\dagger M(\mathbf{Y}_0) (t_1-t_0),
\end{split}
\end{equation}
so, ${\mathbf{Y}}_{III}(t_1) = \mathbf{U}_{III}(t_0) {\mathbf{L}}(t_1) =  \mathbf{U}_{III}(t_0)\mathbf{S}_{III}(t_1) \mathbf{V}_{III}(t_1)^{\dagger}$.
The approximation $\mathbf{Y}(t_1) \approx \mathbf{Y}_{III}(t_1)$ completes the propagation step from $t_0$ to $t_1$. 
The next integration step is initialized, as follows: $\mathbf{U}_0=\mathbf{U}_{III}(t_1)$, $\mathbf{S}_0=\mathbf{S}_{III}(t_1)$, and $\mathbf{V}_0=\mathbf{V}_{III}(t_1)$, such that $\mathbf{Y}_0=\mathbf{Y}(t_1)$. 

\section{TT notation}
\label{sec:notation}
This Appendix introduces the TT notation necessary for the derivation of the TT-KSL equations of motion, including left and right unfoldings and orthogonalizations, partial products, and reconstructions, consistent with the TT literature.\cite{Oseledets2011,holtz2012manifolds,Lubich2015}

\subsection{Tensor unfolding and reconstruction}
The $i$-th unfolding of a tensor ${X} \in \mathbb{C}^{n_1 \times \cdots \times n_d}$ involves reshaping the tensor as a matrix $\mathbf{X}^{(i)} \in \mathbb{C}^{(n_1 n_2 \cdots n_{i}) \times (n_{i+1} \cdots \times n_d)}$, with entries $\mathbf{X}^{(i)}_{jk}$ corresponding to row $j=n_1 n_2 \cdots n_{i}$ and column $k=n_{i+1} \cdots \times n_d$. Tensor reconstruction is the inverse of unfolding, indicated as follows:
\begin{equation}
\label{eq:ten}
    {X} = \text{Ten}_i[\mathbf{X}^{(i)}].
\end{equation}

\subsection{TT core unfoldings}
TT cores \(X_i\in \mathbb{C}^{r_{i-1}\times n_i\times r_i}\) can be matricized in terms of the so-called left or right unfoldings, denoted as \(\mathbf{X}_i^< \in \mathbb{C}^{r_{i-1} n_i\times r_i}\) and \(\mathbf{X}_i^> \in \mathbb{C}^{r_{i} n_i\times r_{i-1}}\), respectively, as follows:
\begin{equation}
\label{eq:unfold}
\mathbf{X}_i^<=\begin{pmatrix}
X_i(:,1,:) \\
\vdots\\
\vdots\\
\vdots\\
\vdots\\
\vdots\\
\vdots\\
\vdots\\
\vdots\\
X_i(:,n_i,:)\\
\end{pmatrix}
=
\begin{pmatrix}
X_i(1,1,1)&\cdots &X_i(1,1,r_i) \\
\vdots&\cdots&\vdots\\
X_i(r_{i-1},1,1)&\cdots &X_i(r_{i-1},1,r_i)\\
X_i(1,2,1)&\cdots &X_i(1,2,r_i)\\
\vdots&\cdots&\vdots\\
X_i(r_{i-1},2,1)&\cdots &X_i(r_{i-1},2,r_i)\\
\vdots&\cdots&\vdots\\
X_i(1,n_i,1)&\cdots &X_i(1,n_i,r_i)\\
\vdots&\cdots&\vdots\\
X_i(r_{i-1},n_i,1)&\cdots& X_i(r_{i-1},n_i,r_i)\\
\end{pmatrix}
\end{equation}
\begin{equation}
\label{eq:unfold2}
\mathbf{X}_i^>=
\begin{pmatrix}
(X_i(:,:,1))^\dagger \\
\vdots\\
\vdots\\
\vdots\\
\vdots\\
\vdots\\
\vdots\\
\vdots\\
\vdots\\
(X_i(:,:,r_i))^\dagger\\
\end{pmatrix}
=\begin{pmatrix}
X_i(1,1,1)^*&\cdots &X_i(r_{i-1},1,1)^* \\
\vdots&\cdots&\vdots\\
X_i(1,n_i,1)^*&\cdots &X_i(r_{i-1},n_i,1)^*\\
X_i(1,1,2)^*&\cdots &X_i(r_{i-1},1,2)^*\\
\vdots&\cdots&\vdots\\
X_i(1,n_i,2)^*&\cdots &X_i(r_{i-1},n_i,2)^*\\
\vdots&\cdots&\vdots\\
X_i(1,1,r_i)^*&\cdots& X_i(r_{i-1},1,r_i)^*\\
\vdots&\cdots&\vdots\\
X_i(1,n_i,r_i)^*&\cdots& X_i(r_{i-1},n_i,r_i)^*\\
\end{pmatrix}
\end{equation}

\subsection{Partial products of TT cores}
\label{sec:partialproduct}
The left partial product \(X_{\leq i}\in \mathbb{C}^{n_1\times ...\times n_i\times r_i}\) is defined, as follows:
\begin{equation}\label{eq:lpp}
X_{\leq i}(l_1,\dots,l_i,a_{i})={X}_1(l_1)\cdots{X}_i(l_i),
\end{equation} 
where $a_i=1,\dots,r_i$ and $l_k=1,\dots,n_k $ for $ k=1,\dots,i$. Analogously, the right partial product \(X_{\geq i+1}\in \mathbb{C}^{r_i\times n_{i+1}\times \cdots\times n_d}\) is defined, as follows: 
\begin{equation}\label{eq:rpp}
X_{\geq {i+1}}(a_i,l_{i+1},\dots,l_d)={X}_{i+1}(l_{i+1})\cdots{X}_d(l_d).
\end{equation} 
Therefore, the unfoldings $\mathbf{X}_{\leq i} \in \mathbb{C}^{(n_1\cdots n_i)\times r_i}$ and $\mathbf{X}_{\geq {i+1}} \in \mathbb{C}^{(n_{i+1}\cdots n_d)\times r_i}$ of the left and right partial products, introduced by Eqs.~(\ref{eq:lpp}) and~(\ref{eq:rpp}), define the $i$-th unfolding of the tensor, as follows:
\begin{equation}
    \mathbf{X}^{\langle{i}\rangle} = \mathbf{X}_{\leq i} \mathbf{X}_{\geq i+1}^\dagger,
    \label{eq:xi}
\end{equation}
and, according to Eq.~(\ref{eq:ten}),
\begin{equation}\label{prod}
X=\text{Ten}_i[\mathbf{X}_{\leq i} \mathbf{X}_{\geq i+1}^\dagger].
\end{equation} 

\subsection{Recursive construction}
\label{sec:recursiveconstruction}
The unfoldings of left and right partial products $\mathbf{X}_{\leq i} \in \mathbb{C}^{(n_1\cdots n_i)\times r_i}$ and $\mathbf{X}_{\geq {i+1}} \in \mathbb{C}^{(n_{i+1}\cdots n_d)\times r_i}$ can be recursively constructed, as follows: 
\begin{equation}\label{unfold}
\begin{split}
 \mathbf{X}_{\leq i}=(\mathbf{X}_{\leq i-1}\otimes \mathbf{I}_{n_{i}})\mathbf{X}_i^<,\\
  \mathbf{X}_{\geq i}=(\mathbf{I}_{n_{i}}\otimes \mathbf{X}_{\geq i+1})\mathbf{X}_i^>,
\end{split}
\end{equation}
for $i=1, \dots, d$, starting with $\mathbf{X}_{\leq 0}=\mathbf{X}_{\geq d+1}=1$, where $\otimes$ is the Kronecker product. Therefore, according to Eqs.~(\ref{eq:xi}) and~(\ref{unfold}), we obtain:
\begin{equation}\label{tunfold}
\begin{split}
\mathbf{X}^{\langle{i}\rangle} =(\mathbf{X}_{\leq i-1}\otimes \mathbf{I}_{n_{i}})\mathbf{X}_i^< \mathbf{X}_{\geq i+1}^\dagger,
\end{split}
\end{equation}
and
\begin{equation}\label{tunfold2}
\begin{split}
\mathbf{X}^{\langle{i-1}\rangle} &=\mathbf{X}_{\leq i-1} ((\mathbf{I}_{n_{i}}\otimes \mathbf{X}_{\geq i+1})\mathbf{X}_i^>)^\dagger,\\
&=\mathbf{X}_{\leq i-1} \mathbf{X}_i^{>{^\dagger}} (\mathbf{I}_{n_{i}}\otimes \mathbf{X}_{\geq i+1})^\dagger.
\end{split}
\end{equation}
\subsection{Left and right orthogonalization}
\label{sec:lnrortho}
The recursive relations, introduced by Eq.~(\ref{unfold}), enable efficient orthogonalization of the left and right partial products by iterative QR decomposition, as follows. 

Starting with $\mathbf{X}_{\leq 2}=(\mathbf{X}_{\leq 1}\otimes \mathbf{I}_{n_{2}})\mathbf{X}_2^<$, we perform a QR decomposition of $\mathbf{X}_{\leq 1} = \mathbf{X}_{1}^< = \mathbf{Q}_{1}^< \mathbf{R}_1$, with $\mathbf{Q}_{1}$ and orthogonal matrix and $R_1$ an upper triangular matrix,  to obtain
\begin{equation}
\begin{split}
    \mathbf{X}_{\leq 2}&=((\mathbf{Q}_{1}^< \mathbf{R}_1) \otimes \mathbf{I}_{n_{2}})\mathbf{X}_2^<,\\
    &=(\mathbf{Q}_{1}^< \otimes \mathbf{I}_{n_{2}}) (\mathbf{R}_{1} \otimes \mathbf{I}_{n_{2}}) \mathbf{X}_2^<.
\end{split}
\end{equation}
Performing a QR decomposition of $(\mathbf{R}_{1} \otimes \mathbf{I}_{n_{2}}) \mathbf{X}_2^<
=\mathbf{Q}_{2}^< \mathbf{R}_2$, we obtain,
\begin{equation}
\begin{split}
    \mathbf{X}_{\leq 2}
    &=(\mathbf{Q}_{1}^< \otimes \mathbf{I}_{n_{2}}) \mathbf{Q}_{2}^< \mathbf{R}_2,\\
\end{split}
\end{equation}
which, according to Eq.~(\ref{eq:lpp}), gives:
\begin{equation}
\begin{split}
    \mathbf{X}_{\leq 2}
    &=\mathbf{Q}_{\leq 2} \mathbf{R}_2.
\end{split}
\end{equation}
Iterating $i$ times, we obtain the left-orthogonalized partial product $\mathbf{X}_{\leq i} = \mathbf{Q}_{\leq i} \mathbf{R}_i$. Analogously, we obtain the right-orthogonalized partial product $\mathbf{X}_{\geq {i+1}} = \mathbf{Q}_{\geq {i+1}} \mathbf{R}_{i+1}$. Therefore, according to Eq.~(\ref{prod}), we obtain:
\begin{equation}\label{prod2}
\begin{split}
X &= \text{Ten}_i[\mathbf{Q}_{\leq i} \mathbf{R}_i \mathbf{R}_{i+1}^\dagger \mathbf{Q}_{\geq {i+1}}^\dagger],\\
\end{split}
\end{equation}
and introducing the following substitutions $\mathbf{U}_{\leq i}=\mathbf{Q}_{\leq i}$, $\mathbf{S}_i=\mathbf{R}_i \mathbf{R}_{i+1}^\dagger$, and $\mathbf{V}_{\geq {i+1}}=\mathbf{Q}_{\geq {i+1}}$, we obtain the SVD-like decomposition in terms of left and right orthogonalized partial products, as follows:
\begin{equation}\label{ttsvd}
\begin{split}
X 
&= \text{Ten}_i[\mathbf{U}_{\leq i} \mathbf{S}_i \mathbf{V}_{\geq {i+1}}^\dagger].\\
\end{split}
\end{equation}
By recursive construction (Appendix ~\ref{sec:recursiveconstruction}), the $i+1^{th}$ unfolding matrix of a left-and-right orthogonalized $X$ can be obtained given $X^{\langle i\rangle}=\mathbf{U}_{\leq i}\mathbf{S}_i\mathbf{V}_{i+1}^\dagger$. First, substituting $\mathbf{V}_{\geq i+1}=(\mathbf{V}_{\geq i+2}\otimes \mathbf{I}_{n_{i+1}})\mathbf{V}_{i+1}^>$ according to Eq.~\eqref{unfold}:
\begin{equation}\label{xiunfold}
\begin{split}
X^{\langle i \rangle}&=\mathbf{U}_{\leq i}\mathbf{S}_i\mathbf{V}_{\geq i+1}^\dagger\\
&=\mathbf{U}_{\leq i}\mathbf{S}_i\mathbf{V}_{i+1}^{>\dagger}(\mathbf{V}_{\geq i+2}\otimes \mathbf{I}_{n_{i+1}})^\dagger.
\end{split}
\end{equation}
Observing that Eq.~\eqref{xiunfold} is Eq.~\eqref{tunfold2} with $i$ in place of $i-1$, $\mathbf{U}_{\leq i}\mathbf{S}_i$ in place of $\mathbf{X}_{\leq i-1}$, $\mathbf{V}_{i+1}^>$ in place of $\mathbf{X}_i^>$, and $\mathbf{V}_{\geq i+2}$ in place of $\mathbf{X}_{\geq i+1}$, the expression for $X^{\langle i+1\rangle}$ from Eq.~\eqref{xiunfold} is obtained by making these changes of variables in Eq.~\eqref{tunfold}:
\begin{equation}\label{xipounfold}
\begin{split}
X^{\langle i+1\rangle}&=(\mathbf{I}_{n_i}\otimes(\mathbf{U}_{\leq i}\mathbf{S}_i))\mathbf{V}_{i+1}^<\mathbf{V}_{\geq i+2}^\dagger\\
&=(\mathbf{I}_{n_i}\otimes\mathbf{U}_{\leq i})(\mathbf{I}_{n_i}\otimes\mathbf{S}_i)\mathbf{V}_{i+1}^<\mathbf{V}_{\geq i+2}^\dagger.
\end{split}
\end{equation}

\section{Tangent Space and Tensor Train Projection}

\subsection{TT tangent space}
The tangent space $T_Y\mathcal{M}_r$ is defined as the complete set of tensor trains that are tangent to the manifold $\mathcal{M}_r$ of tensor trains $Y$ or rank $r$. Analogous to the definition of the tangent space for matrices, introduced by Eq.~(\ref{defTYMr}), the TT tangent space is defined by the direct sum of mutually orthogonal subsets of tangent tensors, as follows: 
\begin{equation}\label{TxMr}
\begin{split}
T_Y\mathcal{M}_r &=\mathcal{V}_1\oplus \mathcal{V}_1 \oplus \dots \oplus \mathcal{V}_d,\\
\end{split}
\end{equation}
with
\begin{equation}\label{TxMr2}
\begin{split}
\mathcal{V}_j
&=\{\text{Ten}_j[(\mathbf{U}_{\leq j-1}\otimes \mathbf{I}_{n_j})\delta \mathbf{C}_j^<\mathbf{V}_{\geq j+1}^\dagger],\hspace{.2cm}\text{where}\hspace{.2cm} \mathbf{U}_{j}^{<\dagger}\delta\mathbf{C}_j^<=\mathbf{0}\}, \hspace{.2cm}\text{with}\hspace{.2cm}j<d,\\
\mathcal{V}_d &=\{\text{Ten}_j[(\mathbf{U}_{\leq d-1}\otimes \mathbf{I}_{n_d})\delta \mathbf{C}_d^<]\}.
\end{split}
\end{equation}
Note that the gauge condition $\mathbf{U}_{j}^{<\dagger}\delta\mathbf{C}_j^<=\mathbf{0}$ ensures that a displacement of the tensor $X=\text{Ten}_j[\mathbf{U}_{\leq j} \mathbf{U}_{\geq j+1}^\dagger]$ along the direction of an element of $\mathcal{V}_j$ preserves the rank since the number of linearly independent columns of $\mathbf{U}_{j}^<$ is preserved.

We show that the subsets $\mathcal{V}_i$ are indeed mutually orthogonal since the inner product of any two tensors $\delta X_j=(\mathbf{U}_{\leq j-1}\otimes \mathbf{I}_{n_j})\delta \mathbf{C}_j^<\mathbf{V}_{\geq j+1}^\dagger$ and $\delta X_k=(\mathbf{U}_{\leq k-1}\otimes \mathbf{I}_{n_k})\delta \mathbf{C}_k^<\mathbf{V}_{\geq k+1}^\dagger$ from different subsets ({\em i.e.}, $\mathcal{V}_j$ and $\mathcal{V}_k$, with $j \neq k$) is equal to zero, as follows:
\begin{equation}
\begin{split}
\langle \delta X_j \vert \delta X_{k} \rangle &= \text{Tr}[((\mathbf{U}_{\leq j-1}\otimes \mathbf{I}_{n_j})\delta \mathbf{C}_j^<\mathbf{V}_{\geq j+1}^\dagger)^\dagger (\mathbf{U}_{\leq k-1}\otimes \mathbf{I}_{n_k})\delta \mathbf{C}_k^<\mathbf{V}_{\geq k+1}^\dagger ],\\
&= \text{Tr}[(\mathbf{U}_{1} \cdots \mathbf{U}_{j-1} \delta \mathbf{C}_j^< \mathbf{V}_{j+1}^\dagger \cdots \mathbf{V}_{d}^\dagger)^\dagger \mathbf{U}_{1} \cdots \mathbf{U}_{k-1} \delta \mathbf{C}_k^< \mathbf{V}_{k+1}^\dagger \cdots \mathbf{V}_{d}^\dagger ],\\
&= \text{Tr}[\mathbf{V}_{d} \cdots \mathbf{V}_{j+1}  \delta\mathbf{C}_j^{<^\dagger} \mathbf{U}_{j-1}^\dagger \cdots \mathbf{U}_{1}^\dagger \mathbf{U}_{1} \cdots \mathbf{U}_{k-1} \delta \mathbf{C}_k^< \mathbf{V}_{k+1}^\dagger \cdots \mathbf{V}_{d}^\dagger ].
\end{split}
\end{equation}
Assuming that $k > j$, and considering that the cores have been orthogonalized by QR decomposition, $\mathbf{U}_{j}^\dagger \mathbf{U}_{j}=I$, we obtain:
\begin{equation}
\label{eq:e104}
\begin{split}
    \langle \delta X_j \vert \delta X_{k} \rangle 
        &= \text{Tr}[\mathbf{V}_{d} \cdots \mathbf{V}_{j+1}  \delta\mathbf{C}_j^{<^\dagger} \mathbf{U}_{j} \cdots \mathbf{U}_{k-1} \delta \mathbf{C}_k^< \mathbf{V}_{k+1}^\dagger \cdots \mathbf{V}_{d}^\dagger]=0,
\end{split}
\end{equation}
since $\delta\mathbf{C}_j^{<^\dagger} \mathbf{U}_{j} =0$. For $j=k$, we obtain:
\begin{equation}
\label{eq:e105}
\begin{split}
    \langle \delta X_j \vert \delta X_{j} \rangle 
        &= \text{Tr}[\mathbf{V}_{d} \cdots \mathbf{V}_{j+1}  \delta\mathbf{C}_j^{<^\dagger} \delta \mathbf{C}_j^< \mathbf{V}_{j+1}^\dagger \cdots \mathbf{V}_{d}^\dagger ],\\
        &= \text{Tr}[\delta\mathbf{C}_j^{<^\dagger} \delta \mathbf{C}_j^<]=\langle \delta C_j \vert \delta C_{j} \rangle,
\end{split}
\end{equation}
where, in the second row of Eq.~(\ref{eq:e105}), we have used the invariance of the trace with respect to cyclic permutations. 
Therefore, any $\delta X$ that belongs to the TT tangent space  can be written as a direct sum of mutually orthogonal tensors, as follows: 
\begin{equation}
\label{eq:dsum}
\delta X=\sum_{j=1}^d \delta X_j.    
\end{equation}

\subsection{TT projection operator}
Now, we show how the representation of the tangent space $T_Y\mathcal{M}_r$ as the direct sum of mutually orthogonal subsets, introduced by Eqs.~(\ref{TxMr}) and~(\ref{TxMr2}), facilitates the derivation of the TT projection operator, introduced by Eq.~(\ref{PXZmain}).

Given an arbitrary tensor $Z\in \mathbb{C}^{n_{1}\times \cdots \times n_d}$ in TT format ({\em e.g.}, $Z=M(Y)$), the projection operator $P_Y$ onto the tangent space at $Y$ ensures that $\langle P_Y(Z)-Z \vert \delta X \rangle = 0$, where $\delta X \in T_Y\mathcal{M}_r$, so
\begin{equation}
    \langle P_Y(Z)\vert \delta X \rangle = \langle Z \vert \delta X \rangle.
\end{equation}
Considering that both $P_Y(Z)$ and $\delta X$ are in the tangent space, they can both be written in terms of orthogonal decompositions (Eq.~(\ref{eq:dsum})), as follows: 
\begin{equation}
P_Y(Z) = \sum_{j=1}^d \delta U_j,
\label{eq:pyz}
\end{equation}
where $\delta U_j \in \mathcal{V}_j$ and $\delta X = \sum_{j=1}^d \delta X_j$. So, for each mutually orthogonal subspace $\mathcal{V}_j$, we obtain:
 \begin{equation}
     \langle \delta U_j  \vert \delta X_j \rangle =  \langle Z \vert \delta X_j \rangle,
     \label{eq:dudxzdx}
\end{equation}
where $\delta X_j \in \mathcal{V}_j$ with $j=1,\dots, d$.

Considering that $\mathbf{X}$ is left-orthogonalized, according to Eqs.~(\ref{eq:rpp}) and~(\ref{unfold}), $\delta \mathbf{X}_{j}^{\langle j \rangle}=(\mathbf{X}_{\leq j-1}\otimes \mathbf{I}_{n_{j}}) \delta \mathbf{C}_j^< \mathbf{X}_{\ge j+1}^\dagger$ and $\delta \mathbf{U}_{j}^{\langle j \rangle}=( \mathbf{X}_{\leq j-1}\otimes \mathbf{I}_{n_{j}}) \delta \mathbf{B}_j^<  \mathbf{X}_{\ge j+1}^\dagger$. Therefore, 
 \begin{equation}
 \begin{split}
     \langle \delta U_j  \vert \delta X_j \rangle &=  \text{Tr}[(( \mathbf{X}_{\leq j-1}\otimes \mathbf{I}_{n_{j}}) \delta \mathbf{B}_j^<  \mathbf{X}_{\ge j+1}^\dagger)^\dagger ( \mathbf{X}_{\leq j-1}\otimes \mathbf{I}_{n_{j}}) \delta \mathbf{C}_j^<  \mathbf{X}_{\ge j+1}^\dagger],\\
     &=  \text{Tr}[   \mathbf{X}_{\ge j+1} \delta \mathbf{B}_j^{<\dagger} ( \mathbf{X}_{\leq j-1}\otimes \mathbf{I}_{n_{j}})^\dagger ( \mathbf{X}_{\leq j-1}\otimes \mathbf{I}_{n_{j}}) \delta \mathbf{C}_j^< \mathbf{X}_{\ge j+1}^\dagger],\\
     &=  \text{Tr}[ \mathbf{X}_{\ge j+1}^\dagger  \mathbf{X}_{\ge j+1} \delta \mathbf{B}_j^{<\dagger} \delta \mathbf{C}_j^<],\\
    &=  \langle \delta \mathbf{B}_j^< \mathbf{X}_{\ge j+1}^\dagger  \mathbf{X}_{\ge j+1}  \vert \delta \mathbf{C}_j^< \rangle.\\
     \end{split}
     \label{eq:dudx}
\end{equation}
Furthermore,
\begin{equation}
 \begin{split}
     \langle Z \vert \delta X_j \rangle &=  \text{Tr}[\mathbf{Z}^{\langle j \rangle \dagger} ( \mathbf{X}_{\leq j-1}\otimes \mathbf{I}_{n_{j}}) \delta \mathbf{C}_j^<  \mathbf{X}_{\ge j+1}^\dagger],\\
&=  \text{Tr}[\mathbf{X}_{\ge j+1}^\dagger \mathbf{Z}^{\langle j \rangle \dagger} ( \mathbf{X}_{\leq j-1}\otimes \mathbf{I}_{n_{j}}) \delta \mathbf{C}_j^<],\\
&= \langle ( \mathbf{X}_{\leq j-1}\otimes \mathbf{I}_{n_{j}})^\dagger \mathbf{Z}^{\langle j \rangle} \mathbf{X}_{\ge j+1} \vert  \delta \mathbf{C}_j^<\rangle.
     \end{split}
     \label{eq:zdx}
\end{equation}
Substituting Eqs.~(\ref{eq:dudx}) and~(\ref{eq:zdx}) into Eq.~(\ref{eq:dudxzdx}), we obtain:
\begin{equation}
 \begin{split}
   \langle \delta \mathbf{B}_j^< \mathbf{X}_{\ge j+1}^\dagger  \mathbf{X}_{\ge j+1}  \vert \delta \mathbf{C}_j^< \rangle
&= \langle ( \mathbf{X}_{\leq j-1}\otimes \mathbf{I}_{n_{j}})^\dagger \mathbf{Z}^{\langle j \rangle} \mathbf{X}_{\ge j+1} \vert  \delta \mathbf{C}_j^<\rangle,
     \end{split}
     \label{eq:zc1}
\end{equation}
which implies that $\delta \mathbf{B}_j^< \mathbf{X}_{\ge j+1}^\dagger  \mathbf{X}_{\ge j+1} - ( \mathbf{X}_{\leq j-1}\otimes \mathbf{I}_{n_{j}})^\dagger \mathbf{Z}^{\langle j \rangle} \mathbf{X}_{\ge j+1}$ is orthogonal to $\delta \mathbf{C}_j^<$. Considering 
the gauge condition $\mathbf{U}_j^\dagger \delta\mathbf{C}_j^<=0$,
the projection operator $\mathbf{P}_{j}^<= \mathbf{U}_j^< \mathbf{U}_j^{< \dagger}$ onto the range of $\mathbf{U}_j^<$ for $j=1, \dots, d-1$, and $\mathbf{P}_{d}^<=0$, we obtain:
\begin{equation}
    (\mathbf{I}_j-\mathbf{P}_{j}^<)(\delta \mathbf{B}_j^< \mathbf{X}_{\ge j+1}^\dagger  \mathbf{X}_{\ge j+1} - ( \mathbf{X}_{\leq j-1}\otimes \mathbf{I}_{n_{j}})^\dagger \mathbf{Z}^{\langle j \rangle} \mathbf{X}_{\ge j+1}) = 0.
    \label{eq:n124}
\end{equation}
Note that Eq.~\eqref{eq:n124} is the tensor-train generalization of Eq.~\eqref{dotu6}.

Rearranging Eq.~\eqref{eq:n124}, we obtain:
\begin{equation}\label{PBM}
(\mathbf{I}_j-\mathbf{P}_j^<)\delta \mathbf{B}_j^< =(\mathbf{I}_j-\mathbf{P}_j^<)(\mathbf{X}_{\leq j-1}\otimes \mathbf{I}_{n_j})^\dagger\mathbf{Z}^{\langle j \rangle}\mathbf{X}_{\geq j+1}(\mathbf{X}_{\ge j+1}^\dagger  \mathbf{X}_{\ge j+1})^{-1}.
\end{equation}

Using the gauge condition $\mathbf{U}_j^{<\dagger}\delta \mathbf{B}_j=\mathbf{0}$ and the projector $\mathbf{P}_{j}^<= \mathbf{U}_j^< \mathbf{U}_j^{< \dagger}$, we can show that $(\mathbf{I}_j-\mathbf{P}_j^<)\delta \mathbf{B}_j^< = \delta \mathbf{B}_j^< $, since
\begin{equation}\label{orthoB}
\begin{split}
(\mathbf{I}_j-\mathbf{P}_j^<)\delta \mathbf{B}_j^<&=\mathbf{I}_j\delta\mathbf{B}_j^<-\mathbf{P}_j^<\delta\mathbf{B}_j^<\\
&=\delta\mathbf{B}_j^<-\mathbf{U}_j^<\mathbf{U}_j^{<\dagger}\delta\mathbf{B}_j^<\\
&=\delta\mathbf{B}_j^<-\mathbf{0}\\
&=\delta \mathbf{B}_j^<.
\end{split}
\end{equation}

Substituting Eq.~\eqref{orthoB} into the left hand side of Eq.~\eqref{PBM}, we obtain:
\begin{equation}\label{delBi}
\delta \mathbf{B}_j^< =(\mathbf{I}_j-\mathbf{P}_j^<)(\mathbf{X}_{\leq j-1}\otimes \mathbf{I}_{n_j})^\dagger\mathbf{Z}^{\langle j \rangle}\mathbf{X}_{\geq j+1} (\mathbf{X}_{\ge j+1}^\dagger  \mathbf{X}_{\ge j+1})^{-1}. 
\end{equation}

Inserting Eq. \eqref{delBi} into $\delta \mathbf{U}_{j}^{\langle j \rangle}=( \mathbf{X}_{\leq j-1}\otimes \mathbf{I}_{n_{j}}) \delta \mathbf{B}_j^<  \mathbf{X}_{\ge j+1}^\dagger$, we obtain:
\begin{equation}\label{genP}
\delta \mathbf{U}_{j}^{\langle j \rangle}=(\mathbf{X}_{\leq j-1}\otimes \mathbf{I}_{n_j})(\mathbf{I}_j-\mathbf{P}_j^<)(\mathbf{X}_{\leq j-1}\otimes \mathbf{I}_{n_j})^\dagger\mathbf{Z}^{\langle j \rangle}\mathbf{X}_{\geq j+1}.(\mathbf{X}_{\ge j+1}^\dagger  \mathbf{X}_{\ge j+1})^{-1} \mathbf{X}_{\geq j+1}^\dagger.
\end{equation}
Introducing the following projectors,
\begin{equation}
\begin{split}\label{orthoP}
&\mathbf{P}_{\leq j-1}=\mathbf{X}_{\leq j-1}\mathbf{X}_{\leq j-1}^\dagger,\\
&\mathbf{P}_{\leq i}=(\mathbf{X}_{\leq j-1}\otimes \mathbf{I}_{n_j})\mathbf{P}_j^<(\mathbf{X}_{\leq j-1}\otimes \mathbf{I}_{n_j})^\dagger,\\
&\mathbf{P}_{\geq j+1}=\mathbf{X}_{\geq j+1} (\mathbf{X}_{\ge j+1}^\dagger  \mathbf{X}_{\ge j+1})^{-1} \mathbf{X}_{\geq j+1}^\dagger,
\end{split}
\end{equation}
we simplify Eq.~\eqref{genP}, as follows: 
\begin{equation}\begin{split}
\delta \mathbf{U}_{j}^{\langle j \rangle} &=(\mathbf{P}_{\leq j-1}\otimes\mathbf{I}_{n_j}-\mathbf{P}_{\leq j})\mathbf{Z}^{\langle j \rangle}\mathbf{P}_{\geq j+1},\; \; \;  j=1,...,d-1\\
\delta \mathbf{U}_{d}^{\langle d \rangle}&=(\mathbf{P}_{\leq d-1}\otimes\mathbf{I}_{n_d})\mathbf{Z}^{\langle d \rangle}.
\end{split}\label{delMiP}
\end{equation}
Note that $\delta \mathbf{U}_{d}^{\langle d \rangle}$ as defined in Eq.~(\ref{delMiP}) has only one term since $\mathbf{P}_d^<=\mathbf{0}$, so $\mathbf{P}_{\leq d}=\mathbf{0}$. 

Considering that according to Eq.~(\ref{eq:pyz}), $P_Y(Z) = \sum_{j=1}^d \delta U_j$, we obtain:
\begin{equation}\label{PXZ}
P_Y(Z)=\sum_{j=1}^{d-1}\text{Ten}_j[(\mathbf{P}_{\leq j-1}\otimes\mathbf{I}_{n_j})\mathbf{Z}^{\langle j \rangle}\mathbf{P}_{\geq j+1}-\mathbf{P}_{\leq j}\mathbf{Z}^{\langle j \rangle}\mathbf{P}_{\geq j+1}]+\text{Ten}_d[(\mathbf{P}_{\leq d-1}\otimes\mathbf{I}_{n_d})\mathbf{Z}^{\langle d \rangle}].
\end{equation}
To simplify the notation, we define: 
\begin{equation}\label{P+-}
\begin{split}
&P_j^+(Z)=\text{Ten}_j[(\mathbf{P}_{\leq j-1}\otimes\mathbf{I}_{n_j})\mathbf{Z}^{\langle j \rangle}\mathbf{P}_{\geq j+1}],\\
&P_j^-(Z)=\text{Ten}_j[\mathbf{P}_{\leq j}\mathbf{Z}^{\langle i \rangle}\mathbf{P}_{\geq j+1}],\\
\end{split}
\end{equation}
to obtain
\begin{equation}\label{simplenotation}
P_Y(Z)=P_1^+(Z)-P_1^-(Z)+P_2^+(Z)-P_2^-(Z)+\cdots -P_{d-1}^-(Z)+P_d^+(Z). 
\end{equation}

\section{TT-KSL Integration}
\label{sec:ttksl}
\subsection{TT-KSL equations of motion}
\label{sec:eomtt}
The TT-KSL method integrates the equations of motion for $Y_i^+(t)$ and $Y_i^-(t)$, introduced by Eq.~\eqref{Trottermain}, sweeping from left to right to update the cores of the tensor $Y(t_0)$.\cite{Lubich2015} 

The initial state $Y_0$ is right orthogonalized at core 1, as follows (Appendix \ref{sec:lnrortho}):
\begin{equation}
\label{eq:y1p}
\begin{split}
[Y_1^+(t_0)]^{\langle 1 \rangle}&=\mathbf{Y}_{\leq 1}(t_0)\mathbf{R}_2^\dagger(t_0)\mathbf{V}_{\geq 2}^\dagger(t_0),\\
&=\mathbf{K}_1^<(t_0)\mathbf{V}_{\geq 2}^\dagger(t_0),
\end{split}
\end{equation}
where $\mathbf{Y}_{\leq 1}(t_0)$ is core 1 of $Y_0$, and $\mathbf{K}_1^<(t)=\mathbf{Y}_{\leq 1}(t)\mathbf{R}_2^\dagger(t_0)$. With the right-orthogonalized $Y_1^+(t_0)$, we solve the first equation in Eq.~\eqref{Trottermain}, written as the unfolding matrix equation:
\begin{equation}\label{intP1+}
[\dot{Y}_1^+(t)]^{\langle 1\rangle}=[P_1^+(MY_1^+(t))]^{\langle 1 \rangle}.
\end{equation}
We integrate Eq.~(\ref{intP1+}) with $\mathbf{V}_{\geq 2}^\dagger$ fixed in time, analogous to the first step of the KSL integration for matrices (Appendix \ref{sec:kintegration}), as follows:
\begin{equation}\label{Y1+lhs}
[\dot{Y}_1^+(t)]^{\langle 1 \rangle}=\dot{\mathbf{K}}_1^<(t)\mathbf{V}_{\geq 2}^\dagger(t_0),
\end{equation}
which gives
\begin{equation}\label{Y1+}
[Y_1^+(t_1)]^{\langle 1 \rangle}=\mathbf{K}_1^<(t_1)\mathbf{V}_{\geq 2}^\dagger(t_0).
\end{equation}
We obtain $\mathbf{K}_1^<(t_1)$ by integrating the equation of motion for $\mathbf{K}_1^<(t)$ ({\em i.e.}, Eq.~(\ref{EOMK1})), which can be obtained, as follows. We substitute Eq.~\eqref{P+-} into Eq.~\eqref{intP1+}, as follows:
\begin{equation}
\label{eq:e128}
\begin{split}
[P_1^+(MY_1^+(t)]^{\langle 1 \rangle}&=\mathbf{P}_{\leq 0}[\mathbf{MY}_1^+(t)]^{\langle 1\rangle}\mathbf{P}_{\geq 2},
\end{split}
\end{equation}
where $\mathbf{P}_{\leq 0}=1$. Next, we substitute the projection operators in Eq.~(\ref{eq:e128}), according to Eq.~\eqref{orthoP}, and we obtain:
\begin{equation}\label{Y1+rhs}
\begin{split}
[P_1^+(MY_1^+(t)]^{\langle 1 \rangle}&=[\mathbf{MY}_1^+(t)]^{\langle 1 \rangle}\mathbf{V}_{\geq 2}(\mathbf{V}_{\geq 2}^\dagger\mathbf{V}_{\geq 2})^{-1}\mathbf{V}_{\geq 2}^\dagger,\\
&=[\mathbf{MY}_1^+(t)]^{\langle 1 \rangle}\mathbf{V}_{\geq 2}\mathbf{V}_{\geq 2}^\dagger,
\end{split}
\end{equation}
where the second row is obtained with $\mathbf{V}_{\geq 2}^\dagger\mathbf{V}_{\geq 2}=\mathbf{I}$. Finally, we equate the right-hand-sides of Eqs.~(\ref{intP1+}) and~\eqref{Y1+lhs} and substitute $[P_1^+(MY_1^+(t)]^{\langle 1 \rangle}$ according to Eq.~\eqref{Y1+rhs}, to obtain: 
\begin{equation}\label{EOMK1}
\dot{\mathbf{K}}_1^<(t)=[\mathbf{MY}_1^+(t)]^{\langle 1 \rangle}\mathbf{V}_{\geq 2}(t_0). 
\end{equation}
$\mathbf{K}_1^<(t_1)$ is obtained by integration of Eq.~\eqref{EOMK1} from $t_0$ to $t_1$ and substituted into Eq.~\eqref{Y1+} to obtain $Y_1^+(t_1)$. In general, Eq.~\eqref{EOMK1} can be integrated by the Runge-Kutta method. In applications to model systems where $M$ is time-independent, such as in the integration of Eq.~(\ref{eq:shr1}) for the time-independent Hamiltonian of rhodopsin introduced by Eq.~(\ref{retiH}), Eq.~\eqref{EOMK1} corresponds to a constant-coefficient ordinary differential equation (ODE) that can be integrated by computing the action of a matrix exponential in the Krylov subspace ({\em i.e.}, Eq.~(\ref{KSM}), Sec.~\ref{sec:intmatexp}), as implemented in Expokit.\cite{sidje1998expokit}

Having obtained $Y_1^+(t_1)$, we complete the propagation of core $1$ by integrating $\dot{Y}_1^-=P_1^-(MY_1^-)$ with initial condition $Y_1^-(t_0)=Y_1^+(t_1)$. We start by orthogonalizing the first core of $Y_1^-(t_0)$, according to the QR decomposition $\mathbf{K}_1^<(t_1)=\mathbf{U}_1^<(t_1)\mathbf{R}_1^<(t_1)$, as follows:
\begin{equation}\label{Y1-USV}
\begin{split}
[Y_1^-(t_0)]^{\langle 1\rangle}&=[Y_1^+(t_1)]^{\langle 1 \rangle},\\
&=\mathbf{K}_1^<(t_1)\mathbf{V}_{\geq 2}^\dagger(t_0),\\
&=\mathbf{U}_1^<(t_1)\mathbf{R}_1^<(t_1)\mathbf{V}_{\geq 2}^\dagger(t_0),\\
&=\mathbf{U}_{\leq 1}(t_1)\mathbf{R}_1^<(t_1)\mathbf{V}_{\geq 2}^\dagger(t_0),\\
&=\mathbf{U}_{\leq 1}(t_1)\mathbf{S}_1(t_0)\mathbf{V}_{\geq 2}^\dagger(t_0),
\end{split}
\end{equation}
where we have introduced the substitution $\mathbf{S}_1(t_0)=\mathbf{R}_1^<(t_1)$. The fourth equality comes from the recursive construction relationship (Eq.~\eqref{unfold}), which is trivial for $i=1$, and useful for subsequent substeps. Analogous to the second step of the KSL integration for matrices (Appendix \ref{sec:sintegration}), we evolve
\begin{equation}
    \label{eq:ym}
[{Y}_1^-(t)]^{\langle 1 \rangle} =\mathbf{U}_{\leq 1}(t_1)\mathbf{S}_1(t)\mathbf{V}_{\geq 2}^\dagger(t_0),
\end{equation}
from $t_0$ to $t_1$, by keeping $\mathbf{U}_{\leq 1}(t_1)$ fixed and requiring $\dot{\mathbf{V}}_{\geq 2}=\mathbf{0}$, such that only $\mathbf{S}_1$ is allowed to change over time during the propagation. Therefore, $[{Y}_1^-(t)]^{\langle 1 \rangle}$ evolves, as follows:
\begin{equation}\label{P1-Ydot}
[\dot{Y}_1^-(t)]^{\langle 1 \rangle}=\mathbf{U}_{\leq 1}(t_1)\frac{d\mathbf{S}_1(t)}{dt}\mathbf{V}_{\geq 2}^\dagger(t_0).
\end{equation}

We obtain the equation of motion for $\dot{\mathbf{S}}_1(t)$, as follows. We substitute $P_1^-$ according to Eq.~\eqref{P+-}, 
\begin{equation}\label{P-1proj}
\begin{split}
[\dot{Y}_1^-(t)]^{\langle 1 \rangle}&=[P_1^-(MY_1^-(t))]^{\langle 1 \rangle},\\
&=\mathbf{P}_{\leq 1}[MY_1^-(t)]^{\langle 1 \rangle}\mathbf{P}_{\geq 2},\\
&=\mathbf{U}_{\leq 1}\mathbf{U}_{\leq 1}^\dagger[MY_1^-(t)]^{\langle 1 \rangle}\mathbf{V}_{\geq 2}(\mathbf{V}_{\geq 2}^\dagger\mathbf{V}_{\geq 2})^{-1}\mathbf{V}_{\geq 2}^\dagger,\\
&=\mathbf{U}_{\leq 1}\mathbf{U}_{\leq 1}^\dagger[MY_1^-(t)]^{\langle 1 \rangle}\mathbf{V}_{\geq 2}\mathbf{V}_{\geq 2}^\dagger,
\end{split}
\end{equation}
where $\mathbf{U}_{\leq 1}$ and $\mathbf{V}_{\geq 2}$ are kept fixed and $\mathbf{V}_{\geq 2}$ is orthogonalized. Comparing Eqs.~(\ref{P-1proj}) and~(\ref{P1-Ydot}), we obtain the equation of motion for $\mathbf{S}_1(t)$, as follows:
\begin{equation}\label{S1}
\frac{d\mathbf{S}_1(t)}{dt}=\mathbf{U}_{\leq 1}^\dagger(t_1)[MY_1^-(t)]^{\langle 1 \rangle}\mathbf{V}_{\geq 2}(t_0).
\end{equation}
$\mathbf{S}_1(t)$ is obtained by integrating Eq.~(\ref{S1}) with Expokit\cite{sidje1998expokit} and substituted into Eq.~\eqref{eq:ym} to obtain $[Y_1^-(t_1)]^{\langle 1 \rangle}$, as follows:
\begin{equation}\label{Y1i-t}
[Y_1^-(t_1)]^{\langle 1 \rangle}=\mathbf{U}_{\leq 1}(t_1)\mathbf{S}_1(t_1)\mathbf{V}_{\geq 2}^\dagger(t_0).
\end{equation}

Having updated the first core, we proceed with the sweeping method to update the second core according to the next two equations of motion introduced by Eq.~\eqref{Trottermain}, which involve $P_2^+$ and $P_2^-$. The initial state $[Y_2^+(t_0)]^{\langle 2 \rangle} = [Y_1^-(t_1)]^{\langle 2\rangle}$ is obtained by refolding $[Y_1^-(t_1)]^{\langle 1\rangle}$ into $[Y_1^-(t_1)]^{\langle 2 \rangle}$, as follows:
\begin{equation}\label{Y1-refold}
\begin{split}
[Y_2^+(t_0)]^{\langle 2\rangle}&=[\text{Ten}_1[Y_1^-(t_1)]^{\langle 1\rangle}]^{\langle 2 \rangle},\\
&=[\text{Ten}_1[\mathbf{U}_{\leq 1}(t_1)\mathbf{S}_1(t_1)\mathbf{V}_{\geq 2}^\dagger(t_0)]]^{\langle 2 \rangle},\\
&=(\mathbf{I}_{n_1}\otimes \mathbf{U}_{\leq 1}(t_1))(\mathbf{I}_{n_1}\otimes \mathbf{S}_1(t_1))\mathbf{V}_{2}^<(t_0)\mathbf{V}_{\geq 3}^\dagger(t_0), 
\end{split}
\end{equation}
where the last equality used Eqs.~\eqref{xiunfold} and \eqref{xipounfold}. Absorbing $\mathbf{S}_1(t_1)$ into core 2 and defining $\mathbf{K}_2^<(t_0)=(\mathbf{I}_{n_1}\otimes \mathbf{S}_1(t_1))\mathbf{V}_{2}^<(t_0)$, we obtain:
\begin{equation}\label{Y2+t0}
[Y_2^+(t_0)]^{\langle 2\rangle}=(\mathbf{I}_{n_1}\otimes \mathbf{U}_{\leq 1}(t_1))\mathbf{K}_2^<(t_0)\mathbf{V}_{\geq 3}^\dagger(t_0).
\end{equation}
The propagation of Eq.~(\ref{Y2+t0}) follows the same procedure as described for Eq.~(\ref{eq:y1p}) to generate $[Y_2^+(t_1)]^{\langle 2\rangle}$. $[Y_2^-(t_0)]^{\langle 2\rangle}$ is initialized by $[Y_2^+(t_1)]^{\langle 2\rangle}$ and propagated, as described for Eq.~(\ref{Y1-USV}), to generate $[Y_2^-(t_1)]^{\langle 2\rangle}$. The same procedure is sequentially applied to update core-by-core all cores of the tensor train --{\em i.e.}, the so-called `sweeping update' algorithm. 

\subsection{Sweeping algorithm}
The sweeping algorithm\cite{Lubich2015} for updating cores $i=2, \dots,d$, implements the procedure applied for updating the first core introduced in Sec.~\ref{sec:eomtt}, which requires integration of the equation of motion,
$\dot{Y}_i^+(t)=P_i^+(MY_i^+(t))$,
for the time interval $[t_0,t_1]$. The equation is written as the unfolding matrix equation,
\begin{equation}\label{TrotterUF}
 [\dot{Y}_i^+(t)]^{\langle i \rangle}=[P_i^+(MY_i^+(t))]^{\langle i \rangle},
\end{equation}
with initial conditions $Y_i^+(t_0)=Y_{i-1}^-(t_1)$. These quantities are left- and right- orthogonalized in terms of the previously updated core, as described by Eq.~(\ref{Y2+t0}) for $Y_2^+(t_0)$:
\begin{equation}
\begin{split}
[Y_i^+(t_0)]^{\langle i \rangle}&=(\mathbf{U}_{\leq i-1}(t_1)\otimes \mathbf{I}_{n_i})\mathbf{K}_i^<(t_0)\mathbf{V}_{\geq i+1}^\dagger(t_0),
\end{split}
\label{eq:ypi}
\end{equation}
with $\mathbf{V}_{\geq d+1}=1$. 
Eq.~(\ref{eq:ypi}) is true for $i=2$, as shown by Eq.~(\ref{Y2+t0}), and can be shown to be valid for any $i$ by induction.

We update $[Y_i^+(t)]^{\langle i \rangle}$ with constant $\mathbf{U}_{\leq i-1}$ and $\mathbf{V}_{\geq i+1}^\dagger$, by propagating $\mathbf{K}_i^<(t)$ and substituting into Eq.~(\ref{eq:ypi}), as follows:
\begin{equation}\label{YUSV}
\begin{split}
[Y_i^+(t)]^{\langle i \rangle}&=(\mathbf{U}_{\leq i-1}(t_1)\otimes \mathbf{I}_{n_i})\mathbf{K}_i^<(t)\mathbf{V}_{\geq i+1}^\dagger(t_0).
\end{split}
\end{equation}
The equation of motion for $\mathbf{K}_i^<(t)$ is obtained from the explicit time derivative of $[Y_i^+(t)]^{\langle i \rangle}$, as defined by Eq.~\eqref{YUSV}, as follows: 
\begin{equation}\label{EOMlhs}
\begin{split}
[\dot{Y}_i^+(t)]^{\langle i \rangle}&=[P_i^+(MY_i^+)(t)]^{\langle i \rangle},\\
&=\frac{d}{dt}[(\mathbf{U}_{\leq i-1}(t_1)\otimes\mathbf{I})\mathbf{K}_i^<(t)\mathbf{V}_{\geq i+1}^\dagger(t_0)],\\
&=(\mathbf{U}_{\leq i-1}(t_1)\otimes\mathbf{I}_{n_i})\frac{d}{dt}[\mathbf{K}_i^<(t)]\mathbf{V}_{\geq i+1}^\dagger(t_0).
\end{split}
\end{equation}
where, according to Eq.~\eqref{P+-}, 
\begin{equation}\label{EOMrhs0}
\begin{split}
[P_i^+(MY_i^+)(t)]^{\langle i \rangle}&=(\mathbf{P}_{\leq i-1}\otimes\mathbf{I}_{n_i})[\mathbf{M}\mathbf{Y}_i^+(t)]^{\langle i \rangle}\mathbf{P}_{\geq i+1}.\\
\end{split}
\end{equation}
Substituting the projection operators in Eq.~(\ref{EOMrhs0}), according to Eq.~\eqref{orthoP}, we obtain:
\begin{equation}\label{EOMrhs}
\begin{split}
[P_i^+(MY_i^+)(t)]^{\langle i \rangle}
&=((\mathbf{U}_{\leq i-1}(t_1)\mathbf{U}_{\leq i-1}^\dagger(t_1))\otimes\mathbf{I}_{n_i})[\mathbf{M}\mathbf{Y}_i^+(t)]^{\langle i \rangle}\mathbf{V}_{\geq i+1}(t_0)\mathbf{V}_{\geq i+1}^\dagger(t_0),\\
&=(\mathbf{U}_{\leq i-1}(t_1)\otimes\mathbf{I}_{n_i})(\mathbf{U}_{\leq i-1}^\dagger(t_1)\otimes\mathbf{I}_{n_i})[\mathbf{M}\mathbf{Y}_i^+(t)]^{\langle i \rangle}\mathbf{V}_{\geq i+1}(t_0)\mathbf{V}_{\geq j+1}^\dagger(t_0).
\end{split}
\end{equation}
and substituting Eq.~\eqref{EOMrhs} into Eq.~\eqref{EOMlhs}, we obtain:
\begin{equation}\label{dotYAs}
\begin{split}
&(\mathbf{U}_{\leq i-1}(t_1)\otimes\mathbf{I}_{n_i})\frac{d}{dt}[\mathbf{K}_i^<(t)]\mathbf{V}_{\geq j+1}^\dagger(t_0)\\
&=(\mathbf{U}_{\leq i-1}(t_1)\otimes\mathbf{I}_{n_i})(\mathbf{U}_{\leq i-1}(t_1)\otimes\mathbf{I}_{n_i})^\dagger[\mathbf{M}\mathbf{Y}_i^+(t)]^{\langle i \rangle}\mathbf{V}_{\geq j+1}(t_0)\mathbf{V}_{\geq j+1}^\dagger(t_0).
\end{split}
\end{equation}


Simplifying Eq.~\eqref{dotYAs}, we obtain:
\begin{equation}\label{Ki}
\begin{split}
\dot{\mathbf{K}}_i^<=(\mathbf{U}_{\leq i-1}^\dagger(t_1)\otimes\mathbf{I}_{n_i})[\mathbf{M}\mathbf{Y}_i^+(t)]^{\langle i \rangle}\mathbf{V}_{\geq i+1}(t_0). 
\end{split}
\end{equation}

Considering that $\mathbf{U}_{\leq i-1}(t)$ and $\mathbf{V}_{\geq i+1}(t)$ are held fixed in time, we can update $Y_i^+$, as defined by Eq.~\eqref{YUSV}, by updating only core $i$ while keeping all other cores unchanged, as follows:
\begin{equation}\label{Yi+UV}
\begin{split}
[Y_i^+(t_1)]^{\langle i \rangle}&=(\mathbf{U}_{\leq i-1}(t_1)\otimes \mathbf{I}_{n_i})\mathbf{K}_i^<(t_1)\mathbf{V}_{\geq i+1}^\dagger(t_0),
\end{split}
\end{equation}
or in matrix product notation,
\begin{equation}\label{Yi+MP}
Y_i^+(j_1,...j_d,t_1)=\mathbf{U}_1(j_1,t_1)...\mathbf{U}_{i-1}(j_{i-1},t_1)\mathbf{K}_i(j_i,t_1)\mathbf{V}_{i+1}(j_{i+1},t_0)...\mathbf{V}_d(j_d,t_0). 
\end{equation}

Having obtained $Y_i^+(t_1)$, we complete the propagation of core $i$ by integrating $\dot{Y}_i^-=P_i^-(MY_i^-)$ with initial condition $Y_i^-(t_0)=Y_i^+(t_1)$. Similar to the process in Eq.~\eqref{Y1-USV}, we orthogonalize the $i^{th}$ core by substituting $\mathbf{K}_i^<(t_1)$ in Eq.~\eqref{Yi+UV} according to the QR decomposition $\mathbf{K}_i^<(t_1)=\mathbf{U}_i^<(t_1)\mathbf{R}_i(t_1)$, as follows:
\begin{equation}\label{Yi-USV}
\begin{split}
[Y_i^-(t_0)]^{\langle i\rangle}&=[Y_i^+(t_1)]^{\langle i \rangle},\\
&=(\mathbf{U}_{\leq i-1}(t_0)\otimes\mathbf{I}_{n_i})\mathbf{K}_i^<(t_1)\mathbf{V}_{\geq i+1}^\dagger(t_0),\\
&=(\mathbf{U}_{\leq i-1}(t_1)\otimes \mathbf{I}_{n_i})\mathbf{U}_i^<(t_1)\mathbf{R}_i(t_1)\mathbf{V}_{\geq i+1}^\dagger(t_0),\\
&=(\mathbf{U}_{\leq i-1}(t_1)\otimes \mathbf{I}_{n_i})\mathbf{U}_i^<(t_1)\mathbf{R}_i(t_1)\mathbf{V}_{\geq i+1}^\dagger(t_0),\\
&=\mathbf{U}_{\leq i}(t_1)\mathbf{R}_i(t_1)\mathbf{V}_{\geq i+1}^\dagger(t_0),\\
&=\mathbf{U}_{\leq i}(t_1)\mathbf{S}_i(t_0)\mathbf{V}_{\geq i+1}^\dagger(t_0),
\end{split}
\end{equation}
where we have introduced the substitution $\mathbf{S}_i(t_0)=\mathbf{R}_i(t_1)$. Since Eq.~\eqref{Yi-USV} is analogous to Eq.~\eqref{Y1i-t} (where $i=1$),  updating $Y_i^-(t)$ requires the same process used to update $Y_1^-(t)$ (Eq.~\eqref{P1-Ydot} to \eqref{Y2+t0}). Analogous to Eq.~\eqref{Y1-refold}, we obtain:
\begin{equation}\label{Yi-UV}
[Y_i^-(t_1)]^{\langle i+1\rangle}=(\mathbf{I}_{n_i}\otimes \mathbf{U}_{\leq i}(t_1))(\mathbf{I}_{n_i}\otimes \mathbf{S}_i(t_1))\mathbf{V}_{i+1}^<(t_0)\mathbf{V}_{\geq i+2}^\dagger(t_0),
\end{equation}
where $\mathbf{S}_i(t_1)$ is obtained by integrating the following equation:
\begin{equation}\label{Si}
\dot{\mathbf{S}}_i(t)=\mathbf{U}_{\leq i}^\dagger[MY_i^-(t)]^{\langle i \rangle}\mathbf{V}_{\geq i+1}.
\end{equation}
Having obtained $\mathbf{K}_{i+1}^<(t_0)=\mathbf{S}_i(t_1))\mathbf{V}_{i+1}^<(t_0)$, we initialize $Y_{i+1}^+$ as in Eq.~\eqref{YUSV}, as follows:

\begin{equation}\label{Yi+USV}
\begin{split}
[Y_{i+1}^+(t)]^{\langle i+1 \rangle}&=(\mathbf{U}_{\leq i}(t_1)\otimes \mathbf{I}_{n_{i+1}})\mathbf{K}_{i+1}^<(t)\mathbf{V}_{\geq i+2}^\dagger(t_0).
\end{split}
\end{equation}
Updating all cores, according to Eqs.~\eqref{YUSV}--\eqref{Yi+USV}, yields $Y_d^+(t_1)$, which approximates the propagated tensor $Y(t_1)=Y_d^+(t_1)$, since according to Eq.~\eqref{Trottermain}, the $d^{th}$ core involves only $Y_d^+$.   

As mentioned in Sec.~\ref{sec:eomtt} for the propagation of the first core, the equations of motion introduced by Eqs.~(\ref{Ki})
 and~(\ref{Si}) can be integrated by the Runge-Kutta method. Furthermore, in applications to model systems where $M$ is time-independent as in the our application to rhodopsin, those equations can be integrated by computing the action of a matrix exponential in the Krylov subspace with Expokit,\cite{sidje1998expokit}
as shown in the Sec.~\ref{sec:intmatexp} (Eq.~(\ref{KSM})).

\subsection{Integration with matrix exponential}
\label{sec:intmatexp}
This section shows that the right-hand side of Eq.~\eqref{Ki}, with a time-independent $M$, defines a constant-coefficient ODE with respect to $\mathbf{K}_i^<(t)$, an equation that can be integrated by computing the action of a matrix exponential. Analogously, we could show that Eq.~\eqref{Si} defines a constant-coefficient ODE with respect to $\mathbf{S}_i(t)$, such that its equation of motion can be integrated analogously.

$MY_i^+(t)$ can be written by using Eqs.~\eqref{Yi+MP} and~\eqref{TTmat}, as follows:
\begin{equation}\label{MYi+}
\begin{split}
MY_i^+(l_1,...,l_d;t)=&\Bigg(\sum_{j_1}\mathbf{M}_1(l_1,j_1)\otimes \mathbf{U}_1(j_1;t_0)\Bigg)...\Bigg(\sum_{j_{i-1}}\mathbf{M}_{i-1}(l_{i-1},j_{i-1})\otimes \mathbf{U}_{i-1}(j_{i-1};t_0)\Bigg)\\
&\Bigg(\sum_{j_i}\mathbf{M}_i(l_i,j_i)\otimes \mathbf{K}_i(j_i;t)\Bigg)\Bigg(\sum_{j_{i+1}}\mathbf{M}_{i+1}(l_{i+1},j_{i+1})\otimes \mathbf{V}_{i+1}(j_{i+1};t_0)\bigg)...\\
&\Bigg(\sum_{j_{d}}\mathbf{M}_{d}(l_{d},j_{d})\otimes \mathbf{U}_{d}(j_{d};t_0)\Bigg),
\end{split}
\end{equation}
which can be written in the unfolding matrix format, as follows:
\begin{equation}\label{MYum}
[\mathbf{M}\mathbf{Y}_i^+(t)]^{\langle i \rangle}=(\mathbf{I}_{n_i}\otimes \bar{\mathbf{U}}_{\leq i-1}(t_0))\overline{\mathbf{K}_i^<}(t)\bar{\mathbf{V}}_{\geq i+1}^\dagger(t_0),
\end{equation}
where $\bar{\mathbf{U}}_{\leq i-1}$ represents the left-unfolding matrices for the first $i-1$ cores and $\bar{\mathbf{V}}_{\geq i+1}$ represents the right-unfolding matrices for the last $d-i$ cores of Eq.~\eqref{MYi+}. $\overline{\mathbf{K}_i^<}(t)$ corresponds to the $i^{th}$ core, and is obtained by operating $\mathbf{M}_i$ on $\mathbf{K}_i^<(t)$ with proper reshaping. 

Inserting Eq.~\eqref{MYum} into Eq.~\eqref{Ki} yields: 
\begin{equation}\label{MYdel}
\begin{split}
\frac{d}{dt}[\mathbf{K}_i^<(t)]&=(\mathbf{I}_{n_i}\otimes \mathbf{U}_{\leq i-1}^\dagger(t_0))(\mathbf{I}_{n_i}\otimes \bar{\mathbf{U}}_{\leq i-1}(t_0))\overline{\mathbf{K}_i^<}(t)\bar{\mathbf{V}}_{\geq i+1}^\dagger(t_0)\mathbf{V}_{\geq i+1}(t_0),
\end{split}
\end{equation}
which has time dependency only for core $i$.
Defining the right hand side of Eq.~\eqref{MYdel} as the action of a matrix operator $\mathbf{W}_i$ on $\mathbf{K}_i^<(t)$, as follows:
\begin{equation}\label{W}
\mathbf{W}_i\mathbf{K}_i^<(t)=(\mathbf{I}_{n_i}\otimes \mathbf{U}_{\leq i-1}^\dagger(t_0))(\mathbf{I}_{n_i}\otimes \bar{\mathbf{U}}_{\leq i-1}(t_0))\overline{\mathbf{K}_i^<}(t)\bar{\mathbf{V}}_{\geq i+1}^\dagger(t_0)\mathbf{V}_{\geq i+1}(t_0),
\end{equation}
we rewrite Eq.~\eqref{Ki}, as follows:
\begin{equation}\label{Kiw}
\begin{split}
\frac{d}{dt}[\mathbf{K}_i^<(t)]&=\mathbf{W}_i[\mathbf{K}_i^<(t)].
\end{split}
\end{equation}
Considering that $\mathbf{W}_i$ is time-independent, Eq.~(\ref{Kiw}) can be formally integrated, as follows:
\begin{equation}\label{KSM2}
\mathbf{K}_i^<(t_1)=e^{(t_1-t_0)\mathbf{W}_i}~\mathbf{K}_i^<(t_0),
\end{equation}
and the matrix exponential can be numerically computed core-by-core by using the Krylov space method, as implemented in the EXPOKIT package.\cite{Sidje1998} 

\section{Codes for quantum dynamics simulations}
The Python codes for TT-SOKSL, TT-KSL, TT-SOFT, and full-grid SOFT simulations of retinal model are available at: https://github.com/NingyiLyu/TTSOKSL. 

\bibliography{refs,Reti_bib}

\newpage
\begin{figure*}
\includegraphics[scale=0.4]{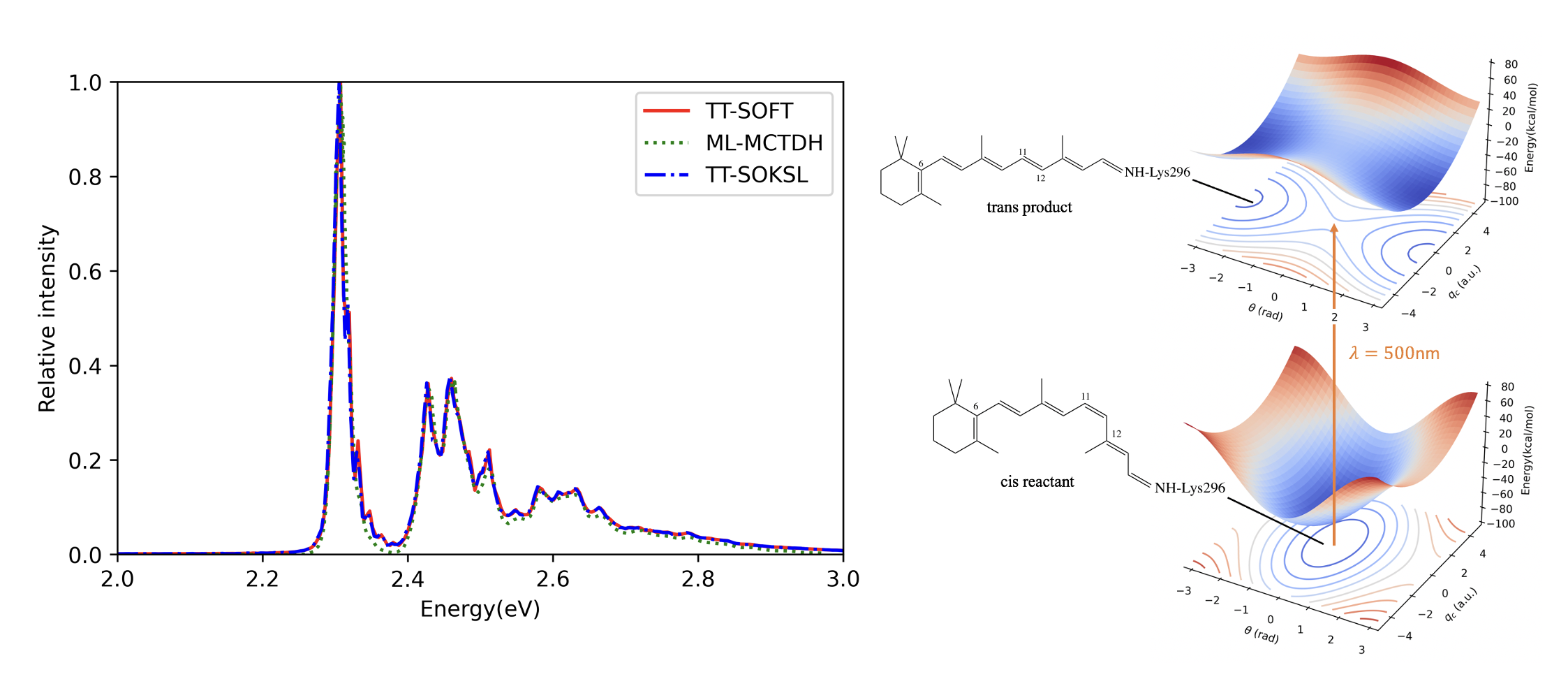}
\caption{Table of contents (TOC).}
\label{fig:TOC}
\end{figure*}

\end{document}